\documentclass[]{spie}  

\usepackage{amsmath,amsfonts,amssymb}
\usepackage{graphicx}
\usepackage[colorlinks=true, allcolors=blue]{hyperref}

\usepackage{float}  

\title{Multiwavelength thermometry without a priori emissivity information: from promise to disillusionment}

\author[a]{Jean-Claude Krapez}
\affil[a]{DOTA, ONERA,
         13300, Salon de Provence, FRANCE}

\authorinfo{E-mail: krapez@onera.fr, Telephone: 33 4 90 17 01 17}

\begin{document} 

\maketitle

\begin{abstract}


Infrared (IR) thermography provides 2D radiance maps of the IR radiation leaving the surfaces of a scene, based on preliminary calibration. Then, to convert radiance maps into temperature maps, we need to know the emissivity of each element of the scene conjugated to each of the detector elements in the camera's focal plane. Compared to the single-band approach, multispectral thermography gives useful insight for detecting anomalies showing distinct spectral signatures, e.g. to detect gas leaks or assess air or water pollution. In addition to that, can spatially distributed multispectral information help solve the inverse temperature-emissivity separation problem? Multiwavelength thermometry (MWT) is known to be an underdetermined problem having a continuous infinity of solutions. For this reason, and right from the origin of MWT, it appeared necessary to introduce information on emissivity to assess temperature, for example, by means of an analytical model. Overlooking these recommendations, a number of papers appeared in the early 2000s exploring the idea that we could do without any a priori knowledge of emissivity. Growth became exponential from 2020 onward, with the allegedly successful application of neural networks, genetic algorithms, and other novel optimization methods.
The aim of the present work is to recall the consequences of MWT as an underdetermined inverse problem, to highlight the errors made by ignoring them, and to bring us back to harsh reality: we \emph{must} introduce information on emissivity in order to evaluate temperature accurately. Furthermore, this information has to be fully consistent with the true spectral emissivity. To this end, we propose a new method for optimal selection of the emissivity model. A series of blind tests was set up to benchmark different inversion algorithms. The results confirm the failure of the temperature-emissivity separation when no emissivity information is available, despite the multiwavelength approach, which in any case comes as no surprise.

v1 is the manuscript of the Keynote paper presented at the conference SPIE 13470, Thermosense: Thermal Infrared Applications XLVII, Orlando USA, 14-17 April 2025.

v2 is a review updated as of June 18, 2025
\end{abstract}

%

\section{INTRODUCTION}
\label{sec:intro}  

\begin{figure}[H]  
\begin{center}
\begin{tabular}{c} 
\includegraphics[height=5cm]{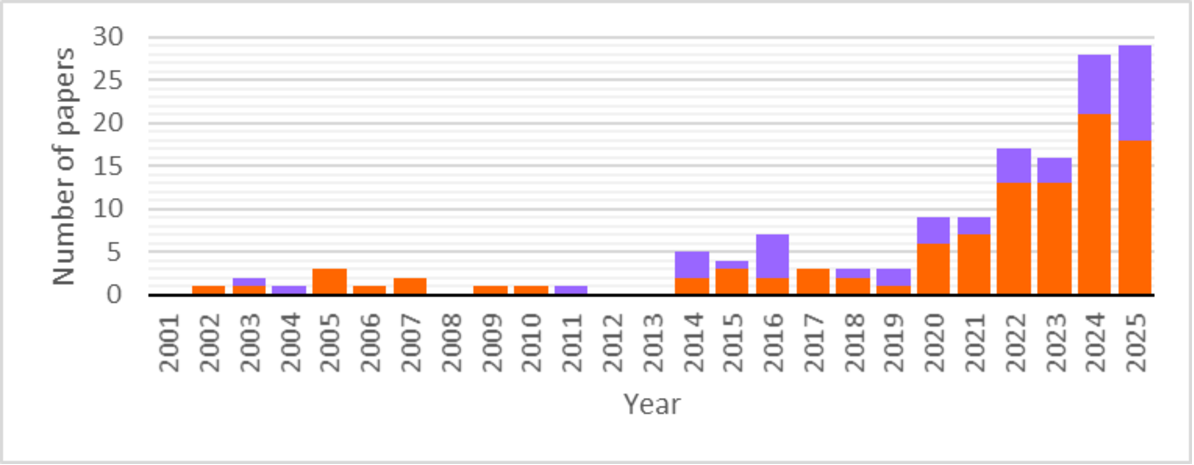}  
\end{tabular}
\end{center}
\caption
{ \label{fig:Nb_of_papers} 
In red: number of papers on multiwavelength thermometry (MWT) claiming that no emissivity information (NEI) is necessary to estimate accurately the temperature, i.e. MWT-NEI methods. In magenta: number of papers on MWT with emissivity models applied blindly (EMAB) (by June 18th 2025).}
\end{figure}

The early 2000s saw the appearance of papers on multiwavelength thermometry (MWT), claiming that the separation of temperature and emissivity could be successfully achieved without introducing any assumptions about emissivity, i.e., without requiring any a priori information about the actual emissivity. For the sake of brevity, this approach, as well as the associated methods and algorithms, will be referred to as MWT-NEI (NEI for 'no emissivity information'), as opposed to MWT-WEI (WEI for 'with emissivity information'). At first, the number of publications on MWT-NEI increased slowly, then, from 2020 onward, an acceleration was observed, as shown in Fig.\ref{fig:Nb_of_papers}. To date, at least 101 papers published in academic journals (i.e., excluding conference papers) have been identified. We here refer to those MWT methods that exclude the necessity of specifying an analytical model for emissivity. We also made a research on MWT methods based on emissivity models but applied \emph{blindly}, we will call them MWT-EMAB, as opposed to MWT-EMAK where the emissivity model is applied \emph{knowingly}. In Fig.\ref{fig:Nb_of_papers} the count of papers on MWT-NEI methods is supposed to be exhaustive, while that on MWT-EMAB is not, by far.


Consider the basic MWT configuration performed at $n$ wavelengths $\lambda_{i}$, $i=1,...,n$, and without reflected radiation. The measured radiance $L_i$ (assumed for the moment to be error free) of the radiation emitted in a given direction and at wavelength $\lambda_{i}$ by a solid surface at temperature $T$ is expressed by:
\begin{equation}
\label{eq:Radiance}
L_i=\epsilon(\lambda_{i},T) B(\lambda_{i},T)\,,\;i=1,...,n
\end{equation}
where $\epsilon(\lambda_{i},T)$ is the surface emissivity at wavelength $\lambda_{i}$ and temperature $T$ and $B(\lambda_{i},T)$ is the blackbody radiance at wavelength $\lambda_{i}$ and temperature $T$. 
The blackbody radiance is given by the Planck's law:
\begin{equation}
\label{eq:Planck}
B(\lambda,T)=\frac{C_1}{\lambda^5}\frac{1}{\exp\left(C_2/(\lambda T)\right)-1},
\end{equation}
or the Wien's approximation:
\begin{equation}
\label{eq:Wien}
B_W(\lambda,T)=\frac{C_1}{\lambda^5}\exp \left(-\frac{C_2}{\lambda T} \right),
\end{equation}
where $C_1$=1.191$\cdot$10$^{-16}$ W~m$^{-2}$ and $C_2$=1.439$\cdot$10$^{-2}$ m~K. The Wien's approximation is better than 1\% provided $\lambda T<$ 3124 µm~K.

It is well known that MWT is an \emph{underdetermined} problem since we have $n+1$ unknowns (the $n$ values of emissivity plus the temperature) but only $n$ measurements of radiance. Basically, the problem has a \emph{multiplicity of solutions}, and for this reason, additional conditions or constraints on emissivity must be introduced (see e.g. Ref.~\citenum{coates1981multi,coates1988least} and the reviews in Ref.~\citenum{krapez2011measurements,
araujo2017multi,
krapez2019measurements}). Coates declared that since the set of equations contains one more unknown than there are equations, it is therefore insoluble "unless some relationship between these can be introduced"~\cite{coates1981multi} or "without at least one additional piece of information"~\cite{coates1988least}.
The constraint on emissivity can take the form of an analytical model for emissivity with a reduced number of parameters as compared to the number of wavelengths. The solution can then be obtained by least-squares optimization. Alternatively, we can introduce a set of a priori values and uncertainties for emissivity and/or temperature and solve the problem in a Bayesian framework. When using an analytical model, it is also well known that it should be chosen very carefully to obtain satisfactory results, otherwise the error on temperature can be very high \cite{krapez2011measurements,
krapez2019measurements}.

Realizing multispectral measurements at several temperature levels, with the aim of having more equations than unknowns, and thus possibly eliminating the ill-posedness of the problem, is just an illusion: with temperature-independent emissivity and applying the Wien approximation, it can be shown that \emph{the problem remains underdetermined}~\cite{kanani2004numerical,krapez2011measurements}. With Planck's formulation, it remains highly ill-conditioned, even in the presence of reflections~\cite{krapez2011measurements,peres2004inverse}.

Because of this burden, i.e. the requirement for a "good" emissivity model, some have been tempted to search for methods that dispense that do without any emissivity information, i.e. MWT-NEI methods. However, from a mathematical logic point of view, this is nonsense, precisely because of the multiplicity of solutions. In fact, these methods and the papers describing them raise a series of questions:
\begin{itemize}
\item how can they manage separating temperature and emissivity despite the fact that the MWT problem is underdetermined and therefore has a multiplicity of solutions ?
\item how can the published results obtained from simulated radiance data be so accurate ?
\item how can the published results obtained from experimental radiance data be so accurate as well ?
\end{itemize}

Should we invoke the presence of informational biases during the inversion tests to explain these last two points ? Information bias is a risk encountered when people try to solve an inverse problem but know the solution in advance.
The best way to avoid this risk is to set up a blind test. This is common practice in many branches of sciences, particularly those related to health - medicine, biology, drug research ...

To this end, a benchmark was set up to assess, through blind tests, the performance of different MWT-NEI methods for the evaluation of temperature and emissivity from multiwavelength radiance data. 
21 invitations to participate in this benchmark have been sent since the beginning of 2024 to research teams having published on this topic over the last 10 years. Spectral radiance data were prepared so that participants could apply their inversion methods.
The radiance data were calculated based on specific configurations considered in most papers dealing with MWT-NEI: single temperature measurement without reflections, single temperature measurement with reflections, multi-temperature measurement without reflections. The underlying question was: which temperature (and which emissivity spectrum) do you think is responsible for the spectral radiance values provided?
Seven responses were received.

This paper will present the results for the following three scenarios studied to evaluate the MWT-NEI methods. They differ from the origin of the supplied radiance data:
\begin{itemize}
\item 1- radiance of purely emitted radiation at one temperature, without reflections, 
\item 2- radiance of emitted plus reflected radiation at one temperature,
\item 3- radiance of emitted radiation when the tested surface is at three different unknown temperature levels while emissivity presents a linear dependence on temperature.
\end{itemize}
Blind test responses will be presented for each case. Then, the \emph{true} temperature and emissivity will be revealed, or, better said, what should be considered the \emph{true} temperature and emissivity. The resulting accuracy of the MWT-NEI methods will be deduced, here without any information bias. This will be followed by a discussion of the different families of MWT-NEI methods and their shortcomings. The demonstrations will make extensive use of a concept that has the virtue of immediately clarifying the nature of the problem and easily revealing the illusory character of the MWT-NEI methods, namely the concept of \emph{permitted} solutions, which was put forward by Coates in Ref.~\citenum{coates1981multi} and later used to advantage in Ref.~\citenum{krapez2011measurements,
krapez2019measurements,rusin2018determination} (Rusin called this tool \emph{relative emissivity}~\cite{rusin2018determination}).
A conclusion will be given on this blind test campaign, which reveals the real potential of MWT-NEI methods and confirms the criticisms expressed in two recent Comments~\cite{krapez2025commentOE,krapez2025commentOL}.

\section{BLIND TESTS FOR BENCHMARKING MWT-NEI METHODS}
\label{sec:Blind_tests}  
A series of blind tests was carried out to assess the performance of different methods for evaluating temperature and emissivity from multiwavelength radiance data.
Three scenarios (S for simple, R for reflections, M for multi-temperature) were considered, with possibly different materials, namely different emissivity spectra~\footnote[1]{in this paper, we call emissivity \emph{spectrum}, the set of $n$ emissivity values corresponding to the $n$ wavelengths of the multispectral measurement, not the continuous set of values taken in the whole spectral band}:
\begin{itemize}
\item S - single temperature, no reflections; we consider only the emitted radiation from material A at temperature $T_{A}$ and material B at temperature $T_{B}$
\item R - single temperature, reflections from the environment considered as a blackbody at known temperature $T_{e}$; we consider the sum of reflected and emitted radiation from material C at temperature $T_{C}$ and material D at temperature $T_{D}$ 
\item M - multi-temperature, no reflections, emissivity is linearly dependent on temperature and the slope is unknown; we consider only the emitted radiation from material  E at three successive temperatures: $T_{1}$, $T_{2}$, and $T_{3}$.
\end{itemize}
For these exercises, either eight wavelengths between 1.4 and 2.5 µm or eleven wavelengths between 1.5 and 2 µm were considered and the spectral radiance leaving the surface was calculated accordingly. The spectral radiance data were then processed according to different inverse methods developed for multispectral thermometry and claimed to require no prior information on emissivity. All but one of the methods were applied blindly by external participants. The method described in Ref. \citenum{zhang2024fast} and called "Euclidean distance optimization" was applied by ourselves for scenario S, based on the software kindly provided by the authors. However, unlike in Ref. \citenum{zhang2024fast}, the constraint on the emissivity search range was relaxed according to what was assumed to be known about emissivity by all other "blind" participants, i.e. nothing. We therefore set the upper limit of the emissivity search range at 1, and the lower limit at 0.02.

\section{BLIND TEST FOR SCENARIO S (SINGLE TEMPERATURE, NO REFLECTIONS)}
\label{sec:Blind_test_S}  
\subsection{Radiance data}
\label{subsec:Radiance_S}
\begin{figure}[htbp]  
\begin{center}
\begin{tabular}{c} 
\includegraphics[height=5cm]{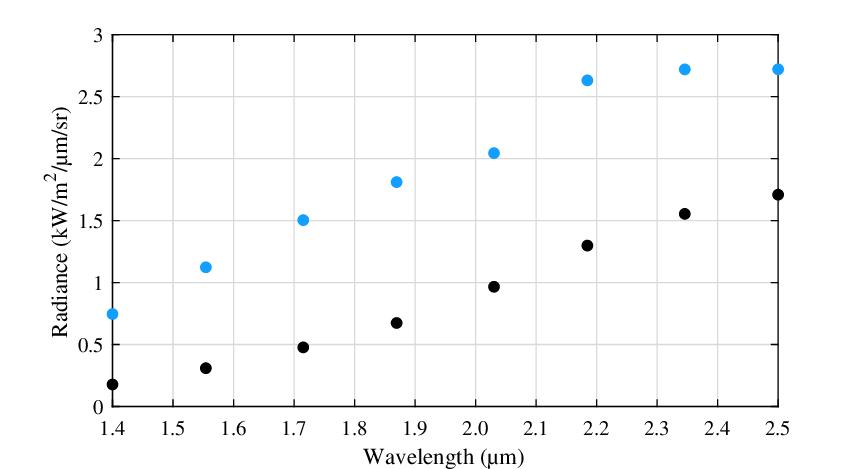}
\end{tabular}
\end{center}
\caption
{\label{fig:Case_S_A_B_Rad} 
Blind test S : Find the temperature of a material knowing the radiance of the emitted radiation shown in the figure. Radiance data for material A are shown in black, and for material B in blue.}
\end{figure}

The radiance emitted by material A and B at eight wavelengths between 1.4 and 2.5 µm is reported in Fig. \ref{fig:Case_S_A_B_Rad}.

\subsection{Inversion results for temperature}
\label{subsec:Inversion_S}
The results of temperature identification are shown in Table \ref{tab:T_Case_S_A} for material A, and in Table \ref{tab:T_Case_S_B} for material B.

\begin{table}[htbp]  
\caption{Temperature inferred from the radiance emitted by material A in scenario S (see Fig. \ref{fig:Case_S_A_B_Rad}) by different inversion methods.} 
\label{tab:T_Case_S_A}
\begin{center}       
\begin{tabular}{|c|c|} 
\hline
\rule[-1ex]{0pt}{3.5ex}  \textbf{Method} & \textbf{Inferred temperature} \\
\hline
\rule[-1ex]{0pt}{3.5ex}  Generalized inverse matrix-graphic deep learning algorithm   \cite{zhang2024multispectralFusion,zhang2023generalized,xing2023graphical}   & 928.6 K\\
\hline
\rule[-1ex]{0pt}{3.5ex}  Euclidean distance optimization \cite{zhang2024fast} & 885.4 K  $^{(1)}$\\
\hline
\rule[-1ex]{0pt}{3.5ex}  Generalized inverse matrix-exterior penalty function \cite{liang2018generalized} & 878.48 K\\
\hline
\rule[-1ex]{0pt}{3.5ex}  Chameleon swarm algorithm  \cite{yao2023chameleon,yao2024multi,he2024computational} & 877.83 K\\
\hline
\rule[-1ex]{0pt}{3.5ex}  Ridge regression \cite{araujo2020surface} & 890 K  $\pm$ 4.35\%\\
\hline
\end{tabular}
\end{center}
\end{table}
$^{(1)}$ : value obtained with an even distribution of emissivity values in the exploratory set for emissivity, as in Ref.~\citenum{zhang2024fast}. Other types of distributions give different temperature values in a very large range.

It should be noted that the Euclidean distance optimization method in Ref.~\citenum{zhang2024fast} is based on a prior selection of emissivity values used as an exploratory set for the \emph{emissivity-way} approach, which is described in more detail in § \ref{subsec:EW}. The distribution of values is freely selectable. The temperature obtained when an even distribution is chosen, as in Ref.~\citenum{zhang2024fast}, is reported in Table \ref{tab:T_Case_S_A} for material A and in Table \ref{tab:T_Case_S_B} for material B. Starting the search with another distribution leads to completely different temperature values. As an example, with a power-law distribution, temperature increases erratically when power decreases. With a power of 0.17, we obtain 1045 K for material A and 1238 K for material B.

\begin{table}[H]  
\caption{Temperature inferred from the radiance emitted by material B in scenario S (see Fig. \ref{fig:Case_S_A_B_Rad}) by different inversion methods.} 
\label{tab:T_Case_S_B}
\begin{center}       
\begin{tabular}{|c|c|} 
\hline
\rule[-1ex]{0pt}{3.5ex}  \textbf{Method} & \textbf{Inferred temperature} \\
\hline
\rule[-1ex]{0pt}{3.5ex}  Generalized inverse matrix-graphic deep learning algorithm \cite{zhang2024multispectralFusion,zhang2023generalized,xing2023graphical}   & 1025.7 K\\
\hline
\rule[-1ex]{0pt}{3.5ex}  Euclidean distance optimization \cite{zhang2024fast} & 1004.4 K  $^{(1)}$\\
\hline
\rule[-1ex]{0pt}{3.5ex}  Generalized inverse matrix-exterior penalty function \cite{liang2018generalized} & 998.78 K\\
\hline
\rule[-1ex]{0pt}{3.5ex}  Chameleon swarm algorithm~ \cite{yao2023chameleon,yao2024multi,he2024computational} & 995.98 K\\
\hline
\rule[-1ex]{0pt}{3.5ex}  Ridge regression \cite{araujo2020surface} & 1011 K  $\pm$ 4.35\%\\
\hline
\rule[-1ex]{0pt}{3.5ex}  Calibration constant constrained secondary measurement method \cite{luo2022emissivity} & 1037.7 K\\
\hline
\end{tabular}
\end{center}
\end{table}
$^{(1)}$ : value obtained with an even distribution of emissivity values in the exploratory set for emissivity, as in Ref.~\citenum{zhang2024fast}. Other types of distributions give different temperature values in a very large range.

\subsection{Comparison with the "true" temperature}
\label{subsec:True_S}
Because MWT is an \emph{underdetermined problem} (which is well known), there is a \emph{continuous infinity of exact solutions} (which is often overlooked). 
These are very easy to obtain via the \emph{temperature-way} originally described by Coates in Ref. \citenum{coates1981multi} and later capitalized in Ref.~\citenum{krapez2011measurements,krapez2019measurements
,rusin2018determination}. We use the term \emph{temperature-way} to refer to this method, as it is based on temperature preselection. Conversely, a large class of MWT-NEI methods is characterized by spectral emissivity preselection, in which case we have introduced the term \emph{emissivity-way}.

In the \emph{temperature-way} approach, trial values, $\hat{T}$, are chosen for temperature, and the related emissivity spectra, $\hat{\epsilon}_{\hat{T},i}$, are obtained from the very simple relation obtained from Equation~(\ref{eq:Radiance}):
\begin{equation}
\label{eq:permitted_emissivity}
\hat{\epsilon}_{\hat{T},i}=L_i/B(\lambda_{i},\hat{T})\,,\;i=1,...,n.
\end{equation}
For any $\hat{T}$ value, and provided that the emissivity values deduced are all within a plausible range, the association of $\hat{T}$ and $\hat{\epsilon}_{\hat{T},i}, i=1,...,n$ constitutes a valid solution to the MWT problem (a \emph{permitted} solution, as named in Ref. \citenum{coates1981multi}), since it explains \emph{perfectly} the measured radiance data, whether error-free or error-corrupted (Rusin, in Ref.~\citenum{rusin2018determination}, used the expression \emph{relative emissivity} to designate a \emph{permitted} spectrum). In this work, we will mainly deal with error-free radiance data. The impact of experimental errors will be discussed in §~\ref{subsec:error}.

\begin{figure}[htbp]  
\begin{center}
\begin{tabular}{c} 
\includegraphics[height=5cm, width=17cm]{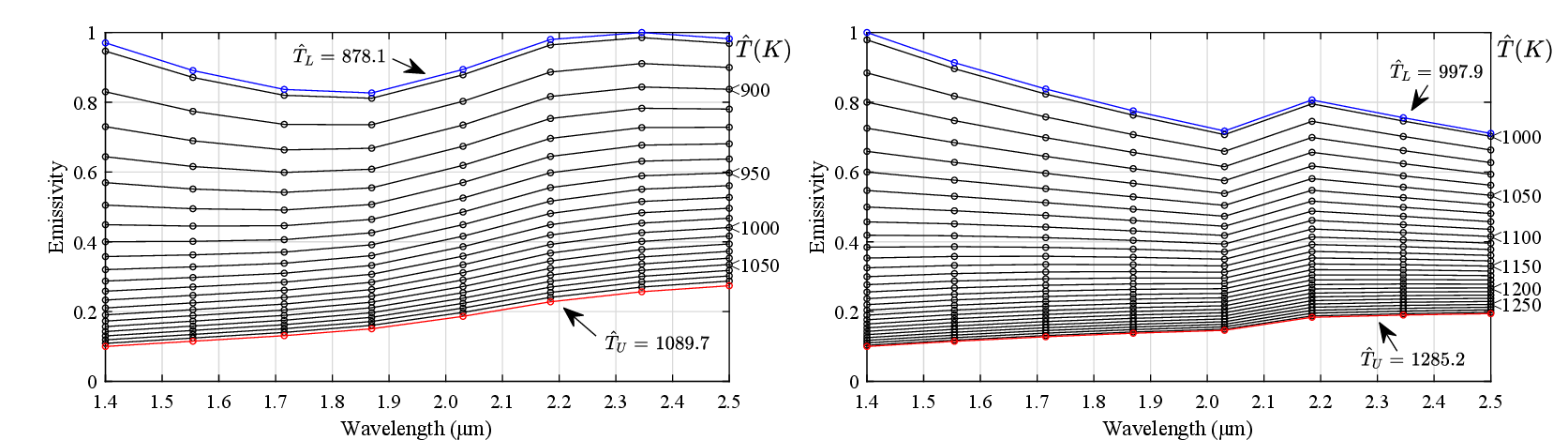}
\end{tabular}
\end{center}
\caption
{ \label{fig:Case_S_A_B_solutions} 
Allowable temperature values $\hat{T}$ and related emissivity spectra $\hat{\epsilon}_{\hat{T},i}$ for scenario S, material A (left) and material B (right). All these combinations of temperature and emissivity lead \emph{exactly} to the respective radiance curves (cases A and B) shown in Fig. \ref{fig:Case_S_A_B_Rad}. Each plot shows a sample of the \emph{infinite} set of \emph{permitted} solutions. For these finite samples, the temperature step for $\hat{T}$ has been set to 10 K. The emissivity spectrum associated to the lower \emph{permitted} temperature, $\hat{T}_{L}$, i.e. such that $max(\hat{\epsilon}_{\hat{T}_{L},i})$=1, is in blue. That associated to the upper \emph{permitted} temperature, $\hat{T}_{U}$, i.e. such that $min(\hat{\epsilon}_{\hat{T}_{U},i})=\epsilon_{min}$, which was here set to 0.1, is in red. The results are represented by dots; the lines between the dots were drawn for ease of reading only.}
\end{figure}

From the radiance data corresponding to material A in Fig. \ref{fig:Case_S_A_B_Rad}, and by applying a 10 K step progression to the exploratory temperature $\hat{T}$ we obtain in Fig. \ref{fig:Case_S_A_B_solutions}-left a subset of the whole set of \emph{permitted} solutions. In the present scenario, the \emph{permitted} temperature and emissivity spectrum evolve in opposite directions. Since the emissivity can be at most 1, we find for the lower limit of the \emph{permitted} temperature, which will be noted $\hat{T}_L$, the value 878.1 K. In theory, there is no upper limit for the \emph{permitted} temperature; indefinitely rising temperature is related to a vanishing emissivity. However, practically, we have to set a plausible value to the lower limit of emissivity. In this work, we chose the value 0.1. As a consequence, the upper limit of the \emph{permitted} temperature, which will be noted $\hat{T}_U$, reaches 1089.7 K.
The results for material B are in Fig. \ref{fig:Case_S_A_B_solutions}-right. In this case, the lower and upper limits of the \emph{permitted} temperature are, respectively $\hat{T}_L$=997.9 K and $\hat{T}_U$=1285.2 K.
Remember that the left and right plots in Fig. \ref{fig:Case_S_A_B_solutions} report only a small subset of the infinite number of \emph{permitted} solutions for materials A and B. In reality, there is a  continuity between the plotted curves.

Interestingly, when ratioing Eq.~(\ref{eq:permitted_emissivity}) for two values of trial temperature $\hat{T}_1$ and $\hat{T}_2$ we obtain~\cite{rusin2018determination}:
\begin{equation}
\label{eq:permitted_emissivity_1_and_2}
\frac{\hat{\epsilon}_{\hat{T}_2,i}}{\hat{\epsilon}_{\hat{T}_1,i}}=\frac{B(\lambda_{i},\hat{T}_1)}{B(\lambda_{i},\hat{T}_2)}\,,\;i=1,...,n ,
\end{equation}
which, with the Wien's approximation, reduces to~\cite{rusin2018determination}:
\begin{equation}
\label{eq:permitted_emissivity_1_and_2_Wien}
\frac{\hat{\epsilon}_{\hat{T}_2,i}}{\hat{\epsilon}_{\hat{T}_1,i}}=\exp\left[\frac{C_2}{\lambda_{i}}\left(\frac{1}{\hat{T}_2}-\frac{1}{\hat{T}_1} \right) \right]
\,,\;i=1,...,n .
\end{equation}
This means that starting from one \emph{permitted} emissivity spectrum, all others can be generated upon multiplication by a simple exponential function $\exp\left(\alpha/\lambda_{i}\right),\;i=1,...,n$, where $\alpha$ is a positive or negative constant.

Now, since all \emph{permitted} solutions obtained by the \emph{temperature-way} approach perfectly explain the observed radiance spectrum reported in Fig. \ref{fig:Case_S_A_B_Rad}, any of them must therefore be considered as a potential and perfectly acceptable candidate for the \emph{true} solution. Indeed, what is important to understand is that the radiance data distributed to the benchmark participants \emph{could have been obtained from any of these combinations} $\left\lbrace \hat{T}, \hat{\epsilon}_{\hat{T},i}, i=1,...,n\right\rbrace$. What's more, there's no argument for rejecting any of these spectra as being too noisy or behaving non-physically, for example. As a consequence, we are allowed to say that \emph{true} solution of the blind test is  \emph{any of the permitted solutions}, a small number of which are shown in Fig. \ref{fig:Case_S_A_B_Rad}. Not only are we allowed to say it, but we are forced to say it for pedagogical purposes. This indeterminacy is indeed a direct consequence of the underdetermination of the MWT problem. This is also a consequence of the fact that the blind test participants didn't ask for information on the true emissivity. Only by using a priori information on true emissivity can the solution space be reduced and the true solution better localized. But it would be contradictory to promote a MWT-NEI method and ask for information on emissivity...

With reference to the above, let's now look at the results of the blind inversions. The methods considered in the bind test provided temperature values in the lower part of the range $[\hat{T}_L, \hat{T}_U$]. In two cases, they are very close to $\hat{T}_L$ (or even very slightly lower, which is not physically permitted). In fact, the values obtained are of little importance in themselves, the important question is: on what basis should either of these temperatures correspond to the true temperature? Actually, as said before, any value between $\hat{T}_L$ and $\hat{T}_U$, together with the corresponding emissivity spectrum reported in Fig. \ref{fig:Case_S_A_B_solutions}, should be considered as possibly the true solution. In the end, any MWT-NEI method gives a result that is statistically as close to the true solution as a random draw between $\hat{T}_L$ and $\hat{T}_U$ would be. Fundamentally, this applies not only to the MWT-NEI methods assessed in the present exercise but to any other MWT-NEI method since, as a matter of principle, they do not leverage any information on emissivity.

\section{BLIND TEST FOR SCENARIO R (SINGLE TEMPERATURE, WITH REFLECTIONS)}
\label{sec:Blind_test_R}  
\subsection{Radiance data}
\label{subsec:Radiance_R}
\begin{figure}[htbp]  
\begin{center}
\begin{tabular}{c} 
\includegraphics[height=5cm]{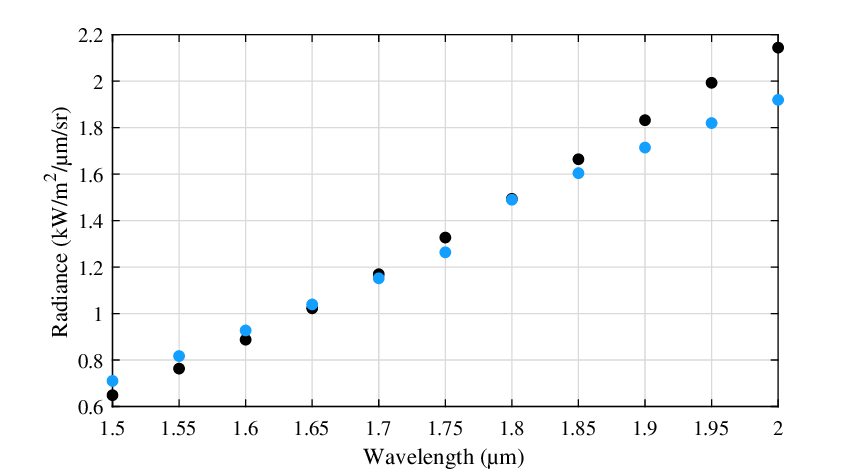}
\end{tabular}
\end{center}
\caption
{ \label{fig:Case_R_C_D_Rad} 
Blind test R : Find the temperature of a material knowing the radiance of the total radiation (emitted+reflected). The environment radiation is assumed to be that of a blackbody at 1000 K. Radiance data for outgoing radiation from material C are shown in black, and for material D in blue.}
\end{figure}

It is assumed that the environment of the surface whose temperature is measured is radiating like a blackbody at 1000 K. This radiation is reflecting on the surface and we collect both the emitted and reflected radiation. The radiance of the radiation leaving the surface of material C and D at eleven wavelengths regularly spaced between 1.5 and 2 µm is reported in Fig. \ref{fig:Case_R_C_D_Rad}. 
\subsection{Inversion results for temperature}
\label{subsec:Inversion_R}
The temperature identification results are in Table \ref{tab:T_Case_R_C_D}. The only method that could be considered for this scenario is a method described in Ref. \citenum{chen2022multi}. It involves Least Squares Support Vector Machine (LSSVM) and Particle Swarm Optimization (PSO), with an emissivity model identification algorithm based on Alpha spectrum-Levenberg
Marquardt (LM) algorithm. The algorithm was able to calculate the temperature for material C but not for material D.
\begin{table}[H]  
\caption{Temperature inferred from the radiance of the radiation leaving material C and D for scenario R (see Fig. \ref{fig:Case_R_C_D_Rad}).} 
\label{tab:T_Case_R_C_D}
\begin{center}       
\begin{tabular}{|c|c|c|}  
\hline
\rule[-1ex]{0pt}{3.5ex}  \textbf{Method} & \textbf{Inferred temperature for mat. C} & \textbf{Inferred temperature for mat. D} \\
\hline
\rule[-1ex]{0pt}{3.5ex}  Alpha-LM method \cite{chen2022multi}   & 949$ \pm$ 0.5 K & ------ \\
\hline
\end{tabular}
\end{center}
\end{table}

\subsection{Comparison with the "true" temperature}
\label{subsec:True_R}
\begin{figure}[htbp]  
\begin{center}
\begin{tabular}{c} 
\includegraphics[height=5cm, width=17cm]{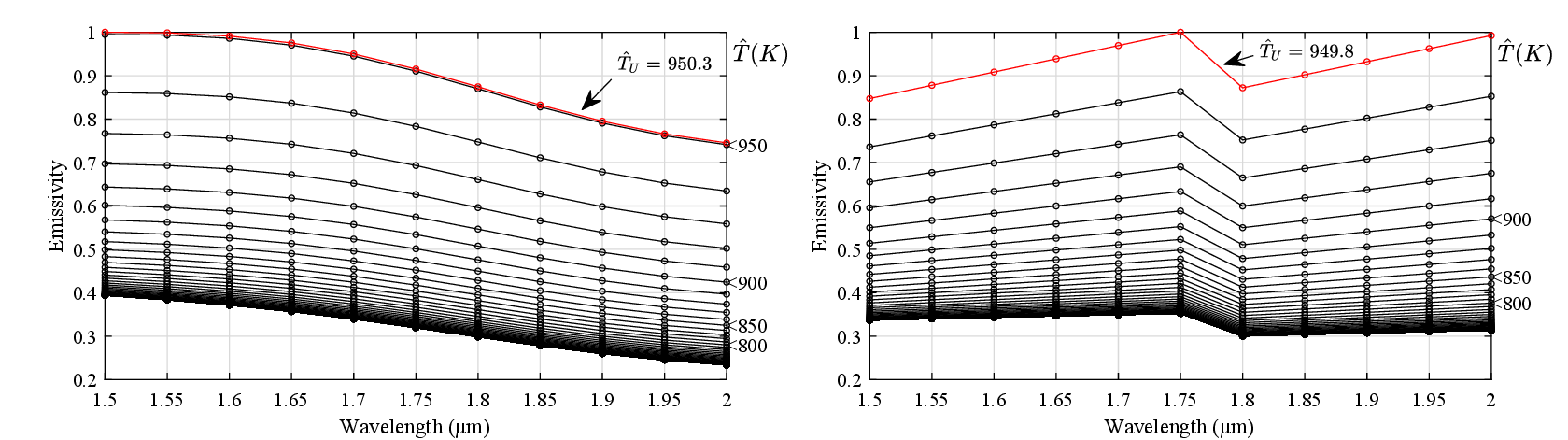}
\end{tabular}
\end{center}
\caption
{ \label{fig:Case_R_C_D_solutions} 
Allowable temperature values $\hat{T}$ and related emissivity spectra $\hat{\epsilon}_{\hat{T},i}$ for scenario R, material C (left) and material D (right). All these combinations of temperature and emissivity lead \emph{exactly} to the respective radiance curves (cases C and D) shown in Fig. \ref{fig:Case_R_C_D_Rad}. Each plot shows a sample of the \emph{infinite} set of \emph{permitted} solutions. For these finite samples, the temperature step for $\hat{T}$ has been set to 10 K. The emissivity spectrum associated to the upper \emph{permitted} temperature, $\hat{T}_{U}$, i.e. such that $max(\hat{\epsilon}_{\hat{T}_{U},i})$=1, is in red. There is not lower limit for the \emph{permitted} temperature. The lower part of the curve bundle represents the asymptotic limit of the spectral emissivity data for a decreasing trial value of temperature. The results are represented by dots; the lines between the dots were drawn for ease of reading only.}
\end{figure}

The presence of reflected radiation from the environment, even in the simple case where this radiation can be modeled by blackbody radiation of known temperature, say $T_e$, does not alleviate the intrinsic difficulty that MWT is an \emph{underdetermined problem} with a \emph{continuous infinity of exact solutions}. 
These can be constructed very easily by adapting the \emph{temperature-way} described by Coates in Ref. \citenum{coates1981multi} to the present situation with a reflection contribution.
Starting from the radiation budget:
\begin{equation}
L(\lambda_{i},T)=\epsilon(\lambda_{i},T) B(\lambda_{i},T)+\left(1-\epsilon(\lambda_{i},T)\right) B(\lambda_{i},T_e)\,,\;i=1,...,n ,
\end{equation}
it is easy to deduce that, from a trial temperature value, $\hat{T}$, the related emissivity spectrum, $\hat{\epsilon}_{\hat{T},i}$, is obtained from the measured radiances $L_i$ according to:
\begin{equation}
\label{eq:eps_refl}
\hat{\epsilon}_{\hat{T},i}=(L_i-B(\lambda_{i},T_e)) / (B(\lambda_{i},\hat{T})-B(\lambda_{i},T_e)), \,,\;i=1,...,n 
\end{equation}
From the radiance data in Fig. \ref{fig:Case_R_C_D_Rad} corresponding to material C, we obtain the subset of \emph{permitted} solutions reported in Fig. \ref{fig:Case_R_C_D_solutions}-left. This subset was again obtained by giving to $\hat{T}$ steps of $10$ K. Unlike the previous scenario, here, the \emph{permitted} temperature and emissivity spectrum evolve in the same direction. This is due to the fact that the temperature of the environment has been set higher than the surface temperature of both material C and D.

Since the emissivity can be at most 1, we find 950.3 K for the upper limit of the \emph{permitted} temperature, $\hat{T}_U$. Here, there is no lower limit for the \emph{permitted} temperature; actually, when decreasing the temperature $\hat{T}$, the related emissivity spectrum approaches an asymptotic limit with a minimum value of 0.234 at 2 µm.

The same general trend can be seen for material D in Fig. \ref{fig:Case_R_C_D_solutions}-right. In this case, the upper limit of the \emph{permitted} temperature, $\hat{T}_U$, is 949.8 K. When lowering the temperature $\hat{T}$, the related emissivity spectrum approaches an asymptotic limit with a minimum value of 0.300 at 1.8 µm.
The Alpha-LM method \cite{chen2022multi} was not able to perform the temperature identification for material D since the presumably sawtooth emissivity spectrum didn't match with the functions that were used by the authors to train the inversion process in a first step. On the other hand, the method gave 949$\pm$0.5 K for the temperature of material C, a value very close to the upper limit of the \emph{permitted} temperature, namely $\hat{T}_U$=950.3 K.

Following the same reasoning as for the previous scenario, here, the "true" temperature can be any value below $\hat{T}_U$, namely 950.3 K for material C and 949.8 K for material D. For a small number of them, the corresponding emissivity spectra are shown in Fig. \ref{fig:Case_R_C_D_solutions}. When no prior information on the real emissivity is available, we cannot go any further. When a MWT-NEI method is claimed to provide the true temperature with accuracy, in fact, this accuracy cannot be better than that obtained by randomly drawing a number less than $\hat{T}_U$. This cannot be otherwise since any of the \emph{permitted} solutions (a small number of which have been reported in Fig. \ref{fig:Case_R_C_D_solutions}) perfectly explains the radiance spectrum of material C, or material D, in Fig. \ref{fig:Case_R_C_D_Rad}. Again, none of these curves can be discarded because of e.g. noisy, erratic, or unphysical variations.
The true solution is hidden among the infinite set of \emph{permitted} solutions and it will remain hidden until prior information on the true emissivity is available.

\begin{figure}[htbp]  
\begin{center}
\begin{tabular}{c} 
\includegraphics[height=5cm, width=17cm]{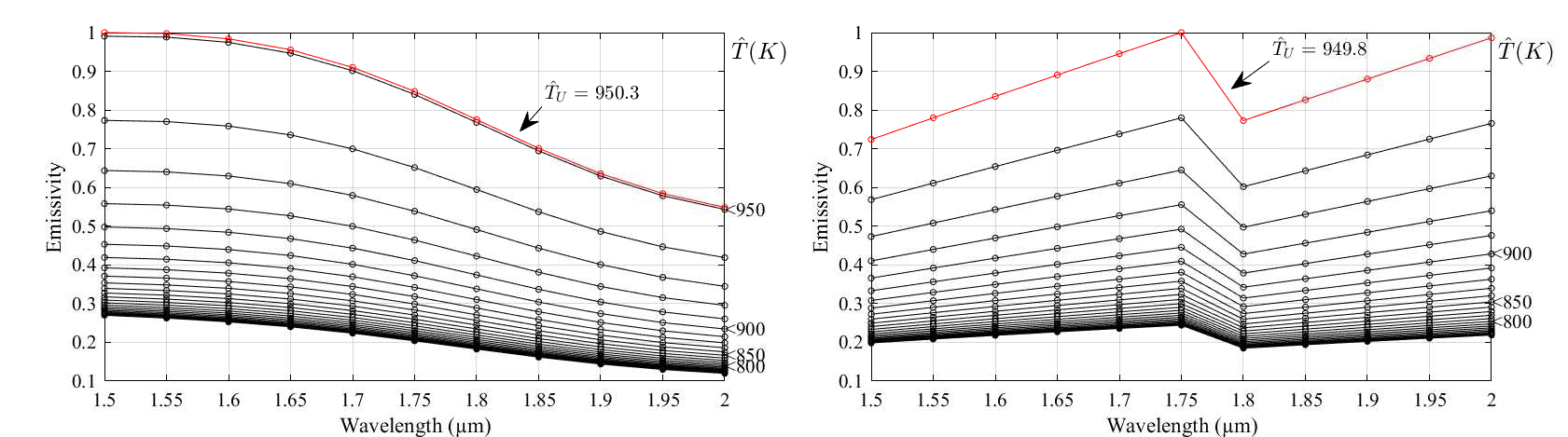}
\end{tabular}
\end{center}
\caption
{ \label{fig:Case_R_C_D_solutions_Te_980} 
Same as in Fig. \ref{fig:Case_R_C_D_solutions} but the environment temperature $T_e$ was here assumed to be 980 K instead of 1000 K.}
\end{figure}

In the present blind test, it was assumed that the radiation from the environment corresponded to that of a blackbody of \emph {known}  temperature $T_e$. If, conversely, this temperature was \emph {unknown}, the difficulty for temperature-emissivity separation would be even worse: the dimension of the solution space would be two instead of one. As an example, we show in Fig. \ref{fig:Case_R_C_D_solutions_Te_980} a finite set of \emph{permitted} solutions obtained when assuming, for the same total radiance values supplied for the blind test, that the environment temperature $T_e$ used in Equation~(\ref{eq:eps_refl}) is not 1000 K (as in Fig. \ref{fig:Case_R_C_D_solutions}) but only 980 K. Unsurprisingly, the \emph{permitted} solutions have changed, even it the range of permitted temperatures has not. This change highlights the fact that the inverse problem is more ill-posed than in the simple case without reflections. Indeed, the underdetermination is then more severe since there are $n$ measured data but as much as $n+2$ unknowns (the $n$ emissivity values, the true temperature and the environment temperature).
To solve this MWT problem correctly, we need information not only on emissivity, but also on environment radiation (assuming it corresponds to blackbody radiation, we need to know the temperature of the environment).

\section{BLIND TEST FOR SCENARIO M (MULTI-TEMPERATURE, NO REFLECTIONS)}
\label{sec:Blind_test_M}  
\subsection{Motivation}
\label{subsec:Motivation_M}
For the reasons explained in the introduction, MWT, when performed at a single temperature, is an \emph{underdetermined problem}. It may be tempting to perform an additional multispectral measurement at another temperature level to solve the underdeterminacy. The reason behind this is that if emissivity does not change with temperature, this brings only one additional unknown (the new temperature value), hence $n+2$ in total, and in the same time, the number of equations increases from $n$ to $2n$. So, as soon as $n\geq2$, the number of equations is greater than or equal to the number of unknowns and we might expect the problem to lose its underdeterminacy and be solvable. Unfortunately, this is an illusion.

Indeed, it has long been known that when emissivity does not vary with temperature, the multi-temperature option is not effective~\cite{kanani2004numerical,krapez2011measurements}.  In fact, when working with the Wien's approximation for the blackbody radiance and linearizing the radiance equations (by taking the logarithm), it can be shown that the linear system remains underdetermined and we still face a continuously infinite number of exact solutions~\cite{kanani2004numerical,krapez2011measurements}. In fact, the $2n\times(n+2)$ linear system is rank deficient of rank $n+1$, hence not allowing the determination of the $n+2$ unknowns. There is again a continuous infinity of exact solutions, as in the single-temperature case. There are easy to construct, following the \emph{temperature-way}. Let us consider the two set of equations describing the multispectral measurements performed at \emph{true} temperature $T_1$ and $T_2$:
\begin{equation}
\label{eq:Radiance1}
L_{i,1}=\epsilon(\lambda_{i}) B_W(\lambda_{i},T_1)\,,\;i=1,...,n
\end{equation}
\begin{equation}
\label{eq:Radiance2}
L_{i,2}=\epsilon(\lambda_{i}) B_W(\lambda_{i},T_2)\,,\;i=1,...,n
\end{equation}
A first remark is that, without any assumption about emissivity, we can  obtain information about temperatures from the radiance measurements anyway. Unfortunately, it is not about $T_1$ and $T_2$ separately but about the difference of the inverses only. Indeed, by taking the ratio of Eqs.~(\ref{eq:Radiance1}) and (\ref{eq:Radiance2}) we obtain:
\begin{equation}
\label{eq:Radiance_ratio}
\frac{1}{T_2}-\frac{1}{T_1}=\frac{\lambda_{i}}{C_2}\ln\frac{L_{i,1}}{L_{i,2}}\,,\;\forall i,i=1,...,n.
\end{equation}
This is valid for any wavelength $\lambda_{i}$, hence in practice, because of experimental errors, it may be advantageous to obtain it by averaging the spectral results:
\begin{equation}
\label{eq:Radiance_ratio_mean}
\frac{1}{T_2}-\frac{1}{T_1}=\frac{1}{n}\sum_{i=1}^{n}
\frac{\lambda_{i}}{C_2}\ln\frac{L_{i,1}}{L_{i,2}}.
\end{equation}
The \emph{temperature-way} procedure is as follows. Trial values are taken for, say, the first temperature, $\hat{T}_1$, and the related emissivity spectra are obtained from Equation~(\ref{eq:permitted_emissivity}), which here becomes:
\begin{equation}
\label{eq:permitted_emissivity_1}
\hat{\epsilon}_{\hat{T}_1,i}=L_{i,1}/B_W(\lambda_{i},\hat{T}_1)\,,\;i...,n.
\end{equation}
The related second temperature is obtained by introducing $\hat{T}_1$ in Equation~(\ref{eq:Radiance_ratio}), which yields:
\begin{equation}
\label{eq:T2}
\hat{T}_2=\left[\frac{1}{\hat{T}_1}+\frac{\lambda_{i}}{C_2}\ln\frac{L_{i,1}}{L_{i,2}}\right]^{-1}\,,\;\forall i,i=1,...,n,
\end{equation}
or alternatively in Equation~(\ref{eq:Radiance_ratio_mean}). The roles of trial temperatures $\hat{T}_1$ and $\hat{T}_2$ can be interchanged.
The solution space is of dimension 1 as for the single-temperature case. Thus operating at more wavelengths or more temperatures does not help for the underdeterminacy as compared to the single-wavelength and single-temperature case: radiation temperature remains an unsolvable problem when no emissivity information is available.

When working with the Planck expression instead, the underdeterminacy in the strict sense disappears, however, the problem remains strongly ill-conditioned \cite{peres2004inverse,krapez2011measurements}. We no longer have an infinite number of exact solutions, but an infinite number of \emph{quasi-solutions}: they explain the observed radiance not exactly but with a very small error, much smaller than the experimental noise commonly encountered~\cite{krapez2011measurements}. Even in the presence of reflections, the problem remains ill-conditioned \cite{peres2004inverse,krapez2011measurements}.

A large number of MWT-NEI algorithms are based on the so-called Secondary Measurement Method (SMM)~ \cite{sun2005processing,sun2009research,
xing2015data,xing2016emissivity,liang2017development,liang2017rules,
sun2020development,sun2020constraints,yang2022study,
luo2022emissivity,wang2023development,luo2024development}, also known as bi-measurement method (BMM). It relies on multi-temperature measurements while assuming that \emph{emissivity varies linearly with temperature}.
 The question that arises is to what extent this variation in emissivity with temperature has an impact on the ill-posed nature of the problem and whether the claimed performance of the SMM can enable reliable and accurate temperature determination.
A specific blind test was implemented to address this issue. It was assumed that MWT was performed on a material E whose surface was successively set at three different temperatures: $T_1$, $T_2$, and $T_3$ for three multispectral measurements. The spectral measurements were made at $n$=8 wavelengths between 1.4 and 2.5 µm. The true emissivity of material E is $\epsilon_{i,j},  i=1,...,n$ at $T_j,  j=1,...,3$ and the three spectra are linked through the following linear relation
\begin{equation}
\label{eq:emiss_lin}
\epsilon_{i,j}=\epsilon_{i,1}\left(1+k\left(T_j-T_1\right)\right)
\,,\;j=2,3, 
\end{equation}
where $k$ is a constant coefficient.

\subsection{Radiance data}
\label{subsec:Radiance_M}
\begin{figure}[htbp]  
\begin{center}
\begin{tabular}{c} 
\includegraphics[height=5cm]{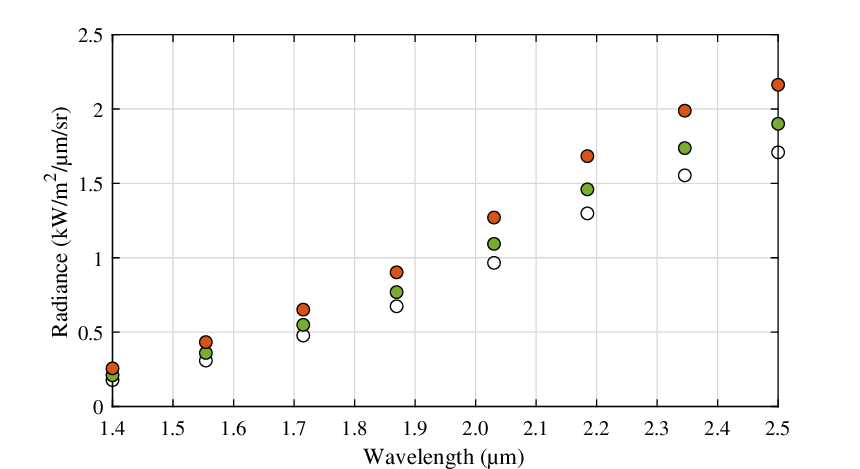}
\end{tabular}
\end{center}
\caption
{ \label{fig:Case_M_1_2_3_Rad} 
Blind test M : Find the temperatures $T_1$, $T_2$, and $T_3$ of a material E of unknown emissivity depending linearly on temperature, knowing the spectral radiance of the radiation emitted at these three different temperature levels. The radiance of the radiation emitted at $T_1$, $T_2$, and $T_3$, is shown with, respectively white, green, and red dots.}
\end{figure}

The radiance leaving the surface of material E at eight wavelengths between 1.4 and 2.5 µm is reported in Fig. \ref{fig:Case_M_1_2_3_Rad}. Data were provided for three temperatures $T_1$, $T_2$, and $T_3$ the participants were asked to estimate.
\subsection{Inversion results for temperature}
\label{subsec:Inversion_M}
Only one MWT-NEI method was blind tested for scenario M: the so-called calibration constant constrained secondary measurement method (CCC-SMM) \cite{luo2022emissivity}. The temperature identification results are in Table \ref{tab:T_Case_M_1_2_3}. 
\begin{table}[H]  
\caption{Temperature inferred from the radiance of the radiation emitted by material E for scenario M at three temperature levels (see Fig. \ref{fig:Case_M_1_2_3_Rad}).} 
\label{tab:T_Case_M_1_2_3}
\begin{center}       
\begin{tabular}{|c|c|c|c|}   
\hline
\rule[-1ex]{0pt}{3.5ex}  \textbf{Method} & \textbf{Estimated $T_1$} & \textbf{Estimated $T_2$} & \textbf{Estimated $T_3$}\\
\hline
\rule[-1ex]{0pt}{3.5ex}  CCC-SMM \cite{luo2022emissivity}   & 927.6 K & 943.9 K  & 963.9 K \\
\hline
\end{tabular}
\end{center}
\end{table}

\subsection{Comparison with the "true" temperatures}
\label{subsec:True_M}
\begin{figure}[htbp]  
\begin{center}
\begin{tabular}{c} 
\includegraphics[height=5cm, width=17cm]{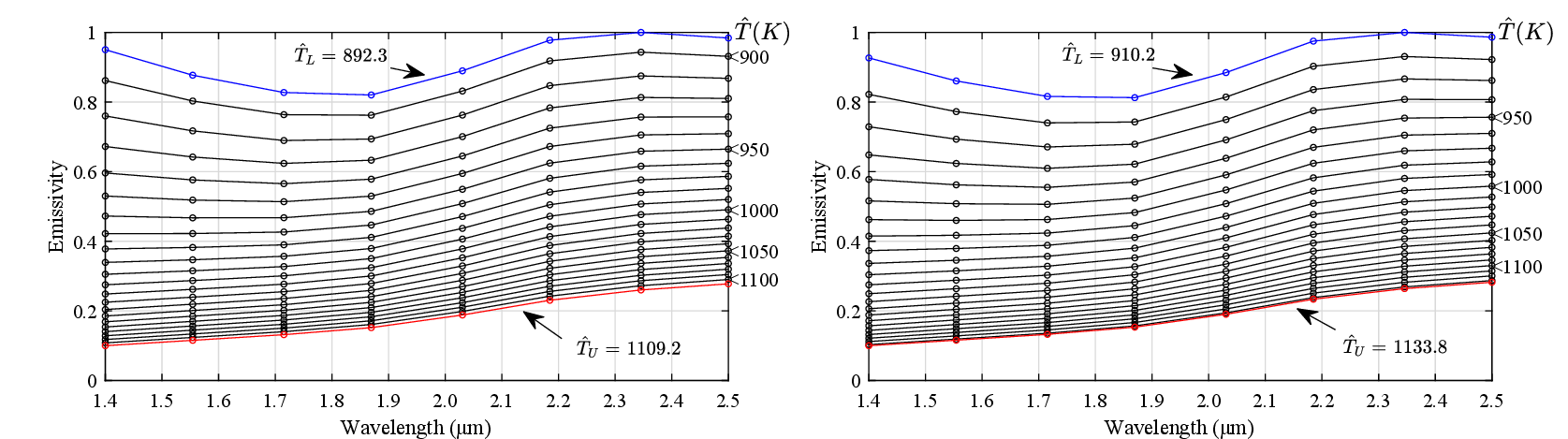}
\end{tabular}
\end{center}
\caption
{ \label{fig:Case_M_2_3_solutions} 
Allowable temperature values $\hat{T}$ and related emissivity spectra $\hat{\epsilon}_{\hat{T},i}$ for scenario M, experience n°2 (left) and experience n°3 (right) when considered separately. All these combinations of temperature and emissivity lead \emph{exactly} to the respective radiance curves (n°2 and n°3) shown in Fig. \ref{fig:Case_M_1_2_3_Rad}. Each plot shows a sample of the \emph{infinite} set of \emph{permitted} solutions. For these finite samples, the temperature step for $\hat{T}$ has been set to 10 K. The emissivity spectrum associated to the lower \emph{permitted} temperature, $\hat{T}_{L}$, i.e. such that $max(\hat{\epsilon}_{\hat{T}_{L},i})$=1, is in blue. That associated to the upper \emph{permitted} temperature, $\hat{T}_{U}$, i.e. such that $min(\hat{\epsilon}_{\hat{T}_{U},i})=\epsilon_{min}$, which was here set to 0.1, is in red. The results are represented by dots; the lines between the dots were drawn for ease of reading only.}
\end{figure}

In a first step of the discussion, let us consider the three multispectral experiments independently and apply the classical \emph{temperature-way} approach to each of them (i.e. the single-temperature version described in \ref{subsec:True_S}). A sample of the \emph{permitted} solutions related to experiment n°1 is actually in Fig. \ref{fig:Case_S_A_B_solutions}-left since the radiance values in Fig. \ref{fig:Case_M_1_2_3_Rad} corresponding to $T_1$ were chosen to be the same as those in Fig. \ref{fig:Case_S_A_B_Rad} for material A. In addition, samples of the \emph{permitted} solutions related to experiment n°2 and 3 were plotted in Fig. \ref{fig:Case_M_2_3_solutions} on the left and right sides respectively.
Following the reasoning already outlined, by selecting any curve from eauch of these three plots, i.e. Fig. \ref{fig:Case_S_A_B_solutions}-left, Fig. \ref{fig:Case_M_2_3_solutions}-left and Fig. \ref{fig:Case_M_2_3_solutions}-right, or more generally any \emph{permitted} solutions from the three corresponding infinite sets, we manage to \emph{perfectly explain} the radiance data reported by the three curves in Fig. \ref{fig:Case_M_1_2_3_Rad}.  Hence, when imposing \emph{no mutual constraint} on the three emissivity spectra recovered, the
\emph{permitted} solution space for the three temperatures has, in $R^3$, the form of a rectangular parallelepiped of sides $T_1\in[878.1, 1089.7]$, $T_2\in[892.3, 1109.2]$ and $T_2\in[910.2, 1133.8]$. In absolute terms, the choice is totally free in this parallelepiped and the corresponding spectral radiances perfectly match with the target radiances in Fig. \ref{fig:Case_M_1_2_3_Rad}. However, by freely choosing the three temperatures in this parallelepiped, the consequence is that the three associated emissivity spectra may be very far apart, which may not be acceptable. We might want these spectra to be close to each other, or to obey a deterministic relation. In this benchmark exercise, the rule of the game was to respect the linearity constraint expressed in Equation~(\ref{eq:emiss_lin}). We will now see how to take it into account.

Let us consider the objective function
\begin{equation}
\label{eq:cost}
J=\sum_{j=1}^{j=3}\sum_{i=1}^{n}\left(1-\hat{\epsilon}_{i,j}B\left(\lambda_{i},\hat{T_j}\right)/L_{i,j}\right)^2,
\end{equation}
where $j=1,..3$ is the index of the multispectral measurement performed at true temperature $T_j$, and where $\hat{\epsilon}_{i,2}$ and $\hat{\epsilon}_{i,2}$ should be expressed in function of $\hat{\epsilon}_{i,1}$ according to Equation~(\ref{eq:emiss_lin}).
The purpose is to minimize the objective function $J$, \emph{for a series of trial values}, $\hat{T}_1$, for the first level. The minimization is carried out by adjusting the following parameters: the two other temperatures $\hat{T}_2$ and $\hat{T}_3$, the $n$ emissivity spectral values at temperature $\hat{T}_1$, and the coefficient $k$ (as said before, the remaining emissivity spectra at $\hat{T}_2$, and $\hat{T}_3$ are then directly obtained by using Equation~(\ref{eq:emiss_lin})). The relative root mean square error on radiance is then given by RMSE=$\sqrt{J/(3n)}$.

\begin{figure}[htbp]  
\begin{center}
\begin{tabular}{c} 
\includegraphics[height=5cm, width=17cm]{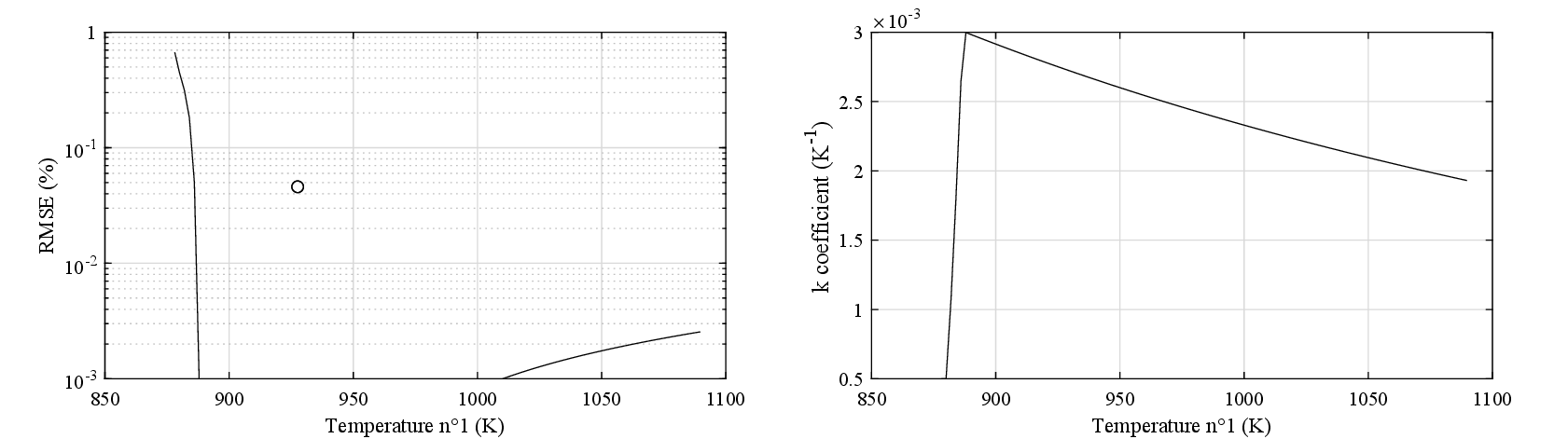}
\end{tabular}
\end{center}
\caption
{ \label{fig:Case_M_1_2_3_RMSE_and_kopt} 
Left: Minimum RMSE (in \%) on radiance as a function of the trial value for the first temperature, $\hat{T}_1$, for the scenario M, by taking all three multispectral measurements into account. The single dot represents the RMS error obtained by the method CCC-SMM \cite{luo2022emissivity} . Right: Optimal value of the linear coefficient $k$ in Equation~(\ref{eq:emiss_lin}) depending on the trial value $\hat{T}_1$.}
\end{figure}

The variations of the minimum RMSE with the trial value $\hat{T}_1$ are plotted in Fig. \ref{fig:Case_M_1_2_3_RMSE_and_kopt}-left. What this curve shows is that in the whole range $[\hat{T}_{1,L}, \hat{T}_{1,U}]$, namely [878 K, 1090 K], the minimum RMSE is lower than 1\%; even more, in the  range [886 K, 1090 K], it is lower than $10^{-2}$\% which could be considered as virtually 0 when compared to the usual level of experimental noise. This means that for a very broad range of the trial temperature $\hat{T}_1$, it is possible to find two other temperatures $\hat{T}_2$ and $\hat{T}_3$ and three compliant emissivity spectra (meaning satisfying the linear equation in  Equation~(\ref{eq:emiss_lin})) so as to almost perfectly match the target radiance values in Fig. \ref{fig:Case_M_1_2_3_Rad}. Given the very low level of the minimum RMSE (in the  range [886 K, 1008 K] for $\hat{T}_1$, it is even lower than $10^{-3}$\%), these \emph{quasi-exact} solutions can be considered as \emph{permitted} solutions  in the same way as those defined in the single-temperature scenarios (scenario S and scenario R).

\begin{figure}[htbp]  
\begin{center}
\begin{tabular}{c} 
\includegraphics[height=5cm, width=17cm]{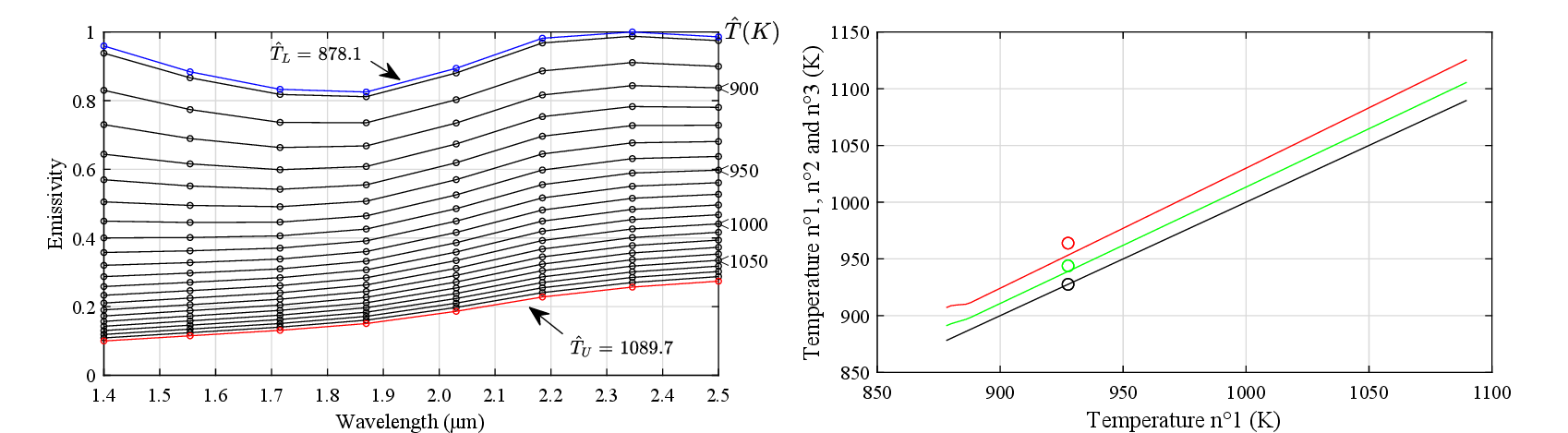}
\end{tabular}
\end{center}
\caption
{ \label{fig:Case_M_1_2_3_solutions_and_T2_T3} 
Left: Allowable temperature values $\hat{T}_1$ and related emissivity spectra at temperature $\hat{T}_1$, i.e. $\hat{\epsilon}_{i,1}$ for scenario M, while taking all tree multispectral measurements into account and upon minimization of the cost function in Equation~(\ref{eq:cost}). Right: relation between the temperature $\hat{T}_1$ on one side and the temperatures $\hat{T}_2$ (in green) and $\hat{T}_3$ (in red) upon minimization of the cost function. The results obtained by the method CCC-SMM \cite{luo2022emissivity} are represented by three colored dots.}
\end{figure} 

The set of \emph{permitted} solutions is characterized as follows, based on the trial temperature $\hat{T}_1$:
\begin{itemize}
\item the optimum value of the linear coefficient $k$ evolves with $\hat{T}_1$ as reported in Fig. \ref{fig:Case_M_1_2_3_RMSE_and_kopt}-right,
\item the relation between the first temperature $\hat{T}_1$ and the corresponding emissivity spectrum is described in Fig. \ref{fig:Case_M_1_2_3_solutions_and_T2_T3}-left. The results are actually very similar to those obtained when considering separately the first multispectral measurement, see Fig. \ref{fig:Case_S_A_B_solutions}-left. The differences appear in the third digit for the lowest temperatures but generally they are only seen in the fifth digit. 
\item the relation between the first temperature $\hat{T}_1$ and the other two temperatures $\hat{T}_2$ and $\hat{T}_3$ is described in Fig. \ref{fig:Case_M_1_2_3_solutions_and_T2_T3}-right. The evolution is closely linear in both cases, except at low temperature, which is due to thresholding of emissivity at 1.
\end{itemize}

The joint values $\hat{T}_1$, $\hat{T}_2$, and $\hat{T}_3$ in Fig. \ref{fig:Case_M_1_2_3_solutions_and_T2_T3}-right constitute quasi-exact solutions. In the 3D space, they form a nearly straight line, which is included in the rectangular parallelepiped discussed previously. There is thus a 1D continuous infinity of \emph{permitted} solutions to the multi-temperature problem. Let us mention, however, that this series of triplets {$\hat{T}_1$, $\hat{T}_2$, $\hat{T}_3$}  was obtained by limiting ourselves to the results of RMSE \emph{minimization}. By relaxing the selection criterion and thus retaining all triplets leading to a RMSE less than a very small value, say $10^{-2}$\%, the solution set would be much wider. In the 3D space, it would take the form an elongated cylinder containing the previous "line" of solutions.

\begin{figure}[htbp]  
\begin{center}
\begin{tabular}{c} 
\includegraphics[height=5cm]{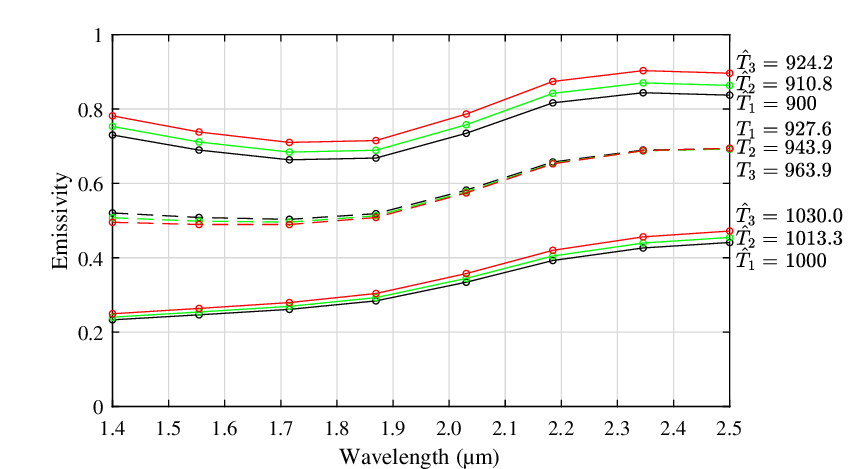}
\end{tabular}
\end{center}
\caption
{\label{fig:Case_M_1_2_3_solution_examples} 
Dots linked with continuous lines represent two examples of \emph{permitted} solutions; the top three curves correspond to the solution generated with $\hat{T}_1$=900 K, the bottom three curves correspond to the solution generated with $\hat{T}_1$=1000 K (the spectrum at $\hat{T}_1$ is in black, that at $\hat{T}_2$ in green, that at $\hat{T}_3$ in red). Dots linked with dashed lines correspond to the solution obtained during the blind test with the CCC-SMM method \cite{luo2022emissivity}. In all cases, the three temperature values (in K) are reported on the right.}
\end{figure}

The inverse method CCC-SMM \cite{luo2022emissivity} provided a \emph{single} triplet of results (see Table \ref{tab:T_Case_M_1_2_3}) that is represented by dots in \ref{fig:Case_M_1_2_3_solutions_and_T2_T3}-right. The corresponding RMSE is 0.046\%. The three emissivity spectra are reported with dashed lines in Fig. \ref{fig:Case_M_1_2_3_solution_examples}.
In Fig. \ref{fig:Case_M_1_2_3_solution_examples} we also reported the spectra corresponding to two particular \emph{permitted} solutions, those generated with $\hat{T}_1$=900 K and $\hat{T}_1$=1000 K. %



The RMSE obtained with the CCC-SMM method was quite low, at 0.046\%, which is deemed acceptable; it can be said that this method succeeded in matching the spectral radiance with the data provided for the blind test. The question, however, is why the three temperatures identified by CCC-SMM should correspond to the truth? We have shown that there is a vast space of \emph{permitted} solutions that explain the radiance data better than $10^{-2}$\%. The true solution could be any other of these \emph{permitted} solutions, since in practice it would be impossible to see the difference when considering the emitted radiance only. Hence, for example, the true solution could be that represented in Fig. \ref{fig:Case_M_1_2_3_solution_examples} with 
$\left\lbrace \hat{T}_1, \hat{T}_2, \hat{T}_3 \right\rbrace $=$ \left\lbrace 900, 910.8, 924.2\right\rbrace$, i.e. the three upper curves or that with $\left\lbrace 1000, 1013.3, 1030\right\rbrace$, i.e. the three lower curves. Both triplets explain the radiance data with a RMSE less than $10^{-3}$\%. What's more, there is a continuous infinity of \emph{permitted} solutions in between.
Again, we have nothing to do but consider one or the other of these as potentially the true solution to assess the success of the blind test. Therefore, again, the accuracy of the solution recovered by the CCC-SMM method is statistically the same as that obtained with a random draw of $\hat{T}_1$ in the range [886 K, 1090 K] followed by the determination of $\hat{T}_1$ and $\hat{T}_1$ from the green, resp. red curves in Fig. \ref{fig:Case_M_1_2_3_solutions_and_T2_T3}-right. This fact holds true for CCC-SMM but also for any other MWT-NEI method that provides a \emph{single} solution claimed to be the true one.

In the end, the single-temperature case and the multi-temperature case present high similarities, leading to the same diagnostic. There is only a slight difference linked to the introduction, for the multi-temperature case, of an acceptance criterion for \emph{quasi-exact} solutions to be considered as \emph{permitted} solutions.
\begin{itemize}
\item on one hand, the literature shows inverse methods that pretend solving the multi-temperature-MWT problem without a priori information on emissivity, except that there may be a linear variation with temperature with unknown coefficient\cite{sun2005processing,sun2009research,
xing2015data,xing2016emissivity,liang2017development,liang2017rules,
sun2020development,sun2020constraints,yang2022study,
luo2022emissivity,wang2023development,luo2024development}. They provide a specific solution that is claimed to correspond (or to be close) to the true solution. The blind test was performed with the 
CCC-SMM method only, but all methods based on SMM share this feature.
\item on the other hand, there is an infinite set of \emph{permitted} solutions that almost perfectly explain the observed spectral radiance (by setting the RMSE to a very low value as for example $10^{-2}$\%, the "almost" perfect solutions can be considered as simply \emph{perfect} solutions). The true solution could be any of these \emph{permitted} solutions; indeed, we wouldn't see any difference on the radiance values, when considering common levels of experimental noise. None of these \emph{permitted} solutions could be ruled out because it is too noisy, or non physical (incidentally, traditional regularization methods applied for overdetermined ill-posed problems would be ineffective, as they are inappropriate here).
\item for all these reasons, the accuracy of the solution recovered by any MWT-NEI method (in particular all methods based on SMM~\cite{sun2005processing,sun2009research,
xing2015data,xing2016emissivity,liang2017development,liang2017rules,
sun2020development,sun2020constraints,yang2022study,
wang2023development,luo2024development}, e.g. CCC-SMM \cite{luo2022emissivity}, is totally unpredictable. That this solution is close to the true solution can only be incidental; random draw from the set of \emph{permitted} solutions would give the same chances of success. The only other possibility for the solution obtained to be accurate is for the range of the true emissivity spectrum to be so wide that the range $[\hat{T}_{L},\,\hat{T}_{U}]$ has shrunk to a value comparable to the accuracy sought. However, in this case, a simple evaluation of these limits, $\hat{T}_{L}$ and $\hat{T}_{U}$, and a random draw in this range would give the true temperature with accuracy with much less effort than the MWT inversion methods. This situation will be referred to as of vanishing \emph{permitted} temperature range (VPTR).
\end{itemize}

One might find it strange to introduce the linearity constraint on emissivity expressed in Equation~(\ref{eq:emiss_lin}). This seems to contradict the fact that MWT-NEI methods are expected not requiring any emissivity information at all. In other words, how can we be sure that the true emissivity follows the linearity constraint whereas nothing else about it is known? In fact, the present work should be considered as a further step of research on temperature-emissivity separation in the scenario of \emph{multi-temperature} measurements:
\begin{itemize}
\item when emissivity does not change with temperature and when using the Wien approximation for the blackbody radiance, the MWT problem remains underdetermined, hence unsolvable;
\item using Planck's expression instead, the MWT problem is not underdetermined but ill-conditioned; it has an infinite number of quasi-exact solutions that can play the classical role of \emph{permitted} solutions, which again makes the problem unsolvable;
\item when the constraint on emissivity is relaxed, namely when it is allowed to depend on temperature, but linearly, we have shown that there is again an infinity of quasi-exact solutions that can play the classical role of \emph{permitted} solutions, hence making the problem unsolvable once again.
\end{itemize}
In all three cases, the problem remains unsolvable if no further information on true emissivity is introduced.

\section{DISCUSSION}
\label{sec:Discussion}
\subsection{Effet of random errors on measurements}
\label{subsec:error}
The influence of the experimental random error will be briefly analyzed in the context of the \emph{temperature-way} approach used to find all \emph{permitted} solutions.
The radiance equation Eq.(\ref{eq:Radiance}) should be rewritten as:
\begin{equation}
\label{eq:Radiance_with_error}
L'_i=\epsilon(\lambda_{i},T) B(\lambda_{i},T)+e_i\,,\;i=1,...,n,
\end{equation}
where $e_i$ is the random error of radiance measurements and $L'_i$ is the noise-corrupted radiance measurement. The \emph{temperature-way} approach keeps the same, hence, after selecting a trial temperature $\hat{T}$, the inferred emissivity spectrum is obtained by applying Eq.~(\ref{eq:permitted_emissivity}) yet with the noise-corrupted radiance at the numerator:
\begin{equation}
\label{eq:permitted_emissivity_with_noise}
\hat{\epsilon}_{\hat{T},i}=L'_i/B(\lambda_{i},\hat{T})\,,\;i=1,...,n .
\end{equation}
This means that the noise corrupting the radiance directly affects the emissivity spectra. Nevertheless, we can say as before that any combination $\hat{T}$ with $\hat{\epsilon}_{\hat{T},i}, i=1,...,n$ constitutes a valid, therefore \emph{permitted} solution to the MWT problem (provided that the emissivity is within a plausible range), since it explains \emph{perfectly} the measured radiance data, even if they are here error-corrupted. The \emph{permitted} emissivity spectra are, of course, noisy, even though the same noise affects identically all of them. It is always possible to apply a filter and recover smoother spectra. But that is not the point, the problem is that as before we have a continuous infinity of temperature solutions. One of them is the true one and we don't know which one until we get information on the true emissivity. Only then can we hope to identify the true temperature, with a degree of uncertainty. Then, the associated \emph{permitted} emissivity spectrum can be presented as a (noisy) estimate of the true emissivity spectrum.
Least squares method, see § \ref{subsec:emissivity_models}, is anyway able to handle noisy data and provide a simultaneous estimate of the temperature and the parameters of the model chosen for the emissivity. Without wishing to spoil what will be said in § \ref{subsec:emissivity_models}, but to avoid any misunderstanding on the interpretation of this “optimal” temperature obtained as a result of the minimization, let's say straight away that it corresponds to the temperature associated with the \emph{permitted} emissivity spectrum that presents the best fitting results (as compared to the other \emph{permitted} emissivity spectra) when using the emissivity model selected for the least-squares method.

\subsection{Flawed approach common to many MWT-NEI methods: the \emph{emissivity-way} and radiance-temperature merging approach}
\label{subsec:EW}
A large number of MWT-NEI methods rely on the \emph{emissivity-way} and radiance-temperature merging approach. It consists first selecting a set of $n$ trial emissivity values $\hat{\epsilon}_{i}, i=1,...,n$, then calculating the corresponding $n$ \emph{radiance temperatures}, $\hat{T}_i$, defined by:
\begin{equation}
\label{eq:T_i}
\hat{T}_i=B^{-1}\left(\lambda_{i},L_i/\hat{\epsilon}_{i}\right)
\,,\;i=1,...,n,
\end{equation}
where $B^{-1}()$ means the blackbody radiance inverse function:
\begin{equation}
\label{eq:B_inverse}
B^{-1}\left(\lambda_{i},L_i\right)=\frac{C_2}{\lambda_{i}\ln \left(1+\frac{C_1}{\lambda_{i}^5 L_i}\right)}.
\end{equation}

In the presence of reflections, the numerator $L_i$ in Equation~(\ref{eq:T_i}) should be replaced by $L_i-(1-\hat{\epsilon}_{i})B(\lambda_i,T_e)$.

If the \emph{true} emissivity spectrum had been selected from the start for the trial values $\hat{\epsilon}_{i}, i=1,...,n$, the radiance temperature values, $\hat{T}_i$ would be reduced to a single value and, of course, this value would correspond to the \emph{true} temperature. However, since the \emph{true} emissivity spectrum is unknown, there is little chance that the inferred $n$ radiance-temperature values are the same. Hence, the strategy of the different inverse methods found in the literature is to progressively adjust the trial emissivity values $\hat{\epsilon}_{i}, i=1,...,n$ until the inferred radiance-temperature values $\hat{T}_i, i=1,...,n$ are acceptably close to each other. This proximity has been quantified using, for example, the following cost function (or objective function):
 \begin{equation}
\label{eq:cost_T_i}
J=\sum_{i=1}^{n}\left(\hat{T}_i-\frac{1}{n}\sum_{j=1}^{j=n}\hat{T}_j\right)^2,
\end{equation}
but many other variants can be found in the literature of \emph{emissivity-way} methods~\cite{yang2003optimum,gao2015analysis,gao2021error,zhang2022multispectral,liang2024two,zhang2024multispectralIdeas ,liu2025data,yang2025data}.
A wide variety of optimization methods have been assessed to efficiently minimize the cost function, thus to \emph{merge the radiance-temperature values}. Many of them were originally developed to solve overdetermined problems.

What motivates the proponents of the \emph{emissivity-way} approach to seek to merge the radiance temperature values is their belief that the mean radiance temperature thus obtained \emph{unconditionally provides an accurate estimate of the true temperature}. This is not correct, as can now be demonstrated:
\begin{itemize}
\item any \emph{permitted} emissivity sprectrum (see e.g. those in Fig. \ref{fig:Case_S_A_B_solutions}), when used in Equation~(\ref{eq:T_i}), leads to a \emph{unique} value of radiance temperature, and this actually corresponds to the trial temperature $\hat{T}$ that was used to generate the \emph{permitted} emissivity sprectrum at stake. The optimization problem that people want to solve by implementing the \emph{emissivity-way} and radiance-temperature merging approach therefore has a continuous infinity of perfect solutions; these are namely nothing else than the \emph{permitted} solutions obtained (by the way, far more easily) with the \emph{temperature-way} approach.
\item that the optimization routine converges to this or that \emph{permitted} solution is a consequence of the internal settings and constraints of the optimization software. As no information on the true emissivity was fed into the optimization processes during the blind tests in §\ref{sec:Blind_test_S}, \ref{sec:Blind_test_R}, \ref{sec:Blind_test_M}, there is no reason for them to converge to the true solution, except incidentally. The same applies to all inversion tests found in the literature related to the \emph{emissivity-way} and radiance-temperature merging approach: since no information on true emissivity is supposed to have been introduced into the optimization processes, there's no reason for them to converge to the true solution, except by chance or in trivial situations where the \emph{permitted} temperature range turns out to be vanishing (these situations will be said to be of vanishing \emph{permitted} temperature range - VPTR). 
Such a situation has already been described, it occurs when the range of true emissivity is so wide that the range of \emph{permitted} temperature has shrunk to such a small value that any temperature value taken therein can be considered accurate anyway. A situation of VPTR can be provoked by setting the range of plausible emissivity values to approximately the same level as that of true emissivity (see Ref.~\citenum{zhang2024fast} and the related comments in  Ref.~\citenum{krapez2025commentOE}).
\end{itemize}

The application of the \emph{emissivity-way} and radiance-temperature merging approach could be traced back to Ref. \citenum{yang2003optimum}. Since then, dozens of papers have been published, conveying the conceptual error that this approach should unequivocally lead to the \emph{true} temperature \cite{yang2003optimum, yang2005measuring,yang2005new,yang2006study,
song2007method,
gao2015multi,gao2015analysis,gao2016multi,
xing2017directly,liang2018generalized,
yu2021multi,wang2021constraint,luan2021light,
gao2021error,zhu2022true,zhang2022ridge,zhang2022multispectral,
tian2022data,shi2022design,chen2022multi,zhao2023approach,
zhang2023data,zhang2023optimizing,yao2023chameleon,
wang2023orthogonal,torres2023radiometric,luan2023temperature,
liang2023multi,
yao2024multi,he2024computational,
zhao2024high,zhang2024multispectralIdeas,zhang2024fast,
zhang2024multispectralWireless,wang2024multi,
wang2024multiReconstructed,tong2024multispectral,
liang2024true,liang2024two,gao2024method,
liu2025data,yang2025data,li2025inverse,gao2025multispectral,
pei2025multi,pei2025high,pei2025multispectral,
pei2025evaluation,
huang2025multispectral,
wang2025advanced}.

In the multi-temperature framework, the motivation behind the effort to merge the radiance-temperature values calculated at the heart of the SMM methods~\cite{sun2005processing,sun2009research,
xing2015data,xing2016emissivity,liang2017development,liang2017rules,
sun2020development,sun2020constraints,yang2022study,
luo2022emissivity,wang2023development,luo2024development} (i.e. separately for each multispectral measurement) suffers from the same conceptual error. Ref.~\citenum{li2025inverse} deals with inversion of infrared thermography data; spatial variations of emissivity are modeled as linear functions of spatial variations of temperature, in line with the SMM approach (quadratic functions have been considered as well).

It is surprising that users of these optimization methods have not observed that not only do they not lead to the true solution, except occasionally, but also that they offer multiple solutions. Indeed, in the hyperspace of parameters and the cost function $J$ (see the expression for it in Equation~(\ref{eq:cost_T_i}) as an example), the values of $J$ form a hypersurface with an extended valley whose bottom is a curve whose level is strictly zero, whether the radiance signal is error-free or not.

A wide variety of methods have been developed to solve the optimization problem posed by the \emph{emissivity-way} and radiance-temperature merging approach, and provide the purported true solution:
\begin{itemize}
\item Neural Networks~\cite{yang2003optimum, yang2005measuring,yang2005new,yang2006study,
song2007method,zhang2024multispectralIdeas}
\item Genetic algorithms~\cite{gao2015multi,gao2015analysis,wang2021constraint,
wang2023orthogonal,zhang2024multispectralIdeas}, in particular different versions of Particle Swarm Optimization (PSO): Least Squares Support Vector Machine (LSSVM) and PSO, with an emissivity model identification algorithm based on Alpha spectrum-Levenberg
Marquardt (LM) algorithm~\cite{chen2022multi}, PSO and Simulated Annealing (SA) algorithms~\cite{zhang2023optimizing}, PSO and the Non-Linear Least Squared method (NLLS)~\cite{torres2023radiometric}, improved fractional-order particle swarm optimization (IFOPSO): high-order nonlinear time-varying inertia weight IFOPSO, and global-local best values IFOPSO~\cite{liang2023multi}, Multi-Strategy PSO~\cite{wang2024multi},
, PSO and the Broyden–Fletcher–Goldfarb–Shanno (BFGS)
algorithm~\cite{tong2024multispectral}, Fractional Order PSO~\cite{liang2024true}, Improved Fractional Order PSO~\cite{liang2024two}, adaptive PSO (APSO) algorithm~\cite{li2025inverse}, Improved
Double-Population Hybrid Genetic Algorithm (IDPHGA)~\cite{liu2025data}
\item Gradient Projection (GP) \cite{xing2017directly,luan2021light} and Internal Penalty Function (IPF)~\cite{xing2017directly,zhu2022true,shi2022design} algorithms
\item BFGS algorithm~\cite{yu2021multi}
\item Sequential Quadratic Programming constraint optimization preceded by ridge regression~\cite{zhang2022ridge}
\item Multi-objective constraint
optimization by mixed
penalty function method~\cite{zhang2022multispectral}
\item Sequential Randomized Coordinate Shrinking (SRCS)
and Multiple-Population Genetic (MPG)~\cite{tian2022data}
\item Hybrid Metaheuristic algorithm based on Multi-Population Genetic (MPG) algorithm and Differential Evolution (DE) algorithm~\cite{zhao2023approach}
\item Combined algorithm of Artificial Bee
Colony and Slime Mould algorithm (CABCSMA) and a Differential Evolution (DE) algorithm~\cite{zhang2023data}
\item Chameleon Swarm algorithm (CSA)~\cite{yao2023chameleon,yao2024multi,he2024computational}
\item moving Emissivity Retardation Spectral Window method (ERSW) based on Lagrange mean value theorem~\cite{luan2023temperature}
\item Improved Grey Wolf Optimization (IGWO)\cite{zhao2024high}
\item Euclidean Distance Optimization~\cite{zhang2024fast,zhang2024multispectralWireless}
\item Barrier Function Interior Point Method (BFIPM)~\cite{yang2025data}
\item Improved Lichtenberg (metaheuristic) algorithm with BP neural network to find emissivity range constraints and Alpha-spectrum model to find emissivity shape constraints~\cite{gao2024method}
\item Barrier Function Interior Point Method (BFIPM)~\cite{yang2025data}
\item Nonlinear inherent emissivity feature extraction method~\cite{gao2025multispectral}
\item Light spectrum Optimizer (LSO), a metaheuristic algorithm stimulating the spectral distibution and peak search in spectral analysis~\cite{wang2025temperature}
\item Generalized Simulated Annealing (GSA) constrained optimization
algorithm~\cite{pei2025multi}, Dual Annealing algorithm~\cite{pei2025high}, Powell Constrained Optimization Algorithm~\cite{pei2025multispectral}, dual Annealing algrithm~\cite{pei2025evaluation}
\item Principal Component Regression (PCR) followed by PSO, then a divide and conquer optimization method integrating the BFGS algorithm~\cite{huang2025multispectral}
\item light spectrum optimizer (LSO) incorporating a Cauchy distribution inverse cumulative function as a mutation factor~\cite{wang2025advanced}
\end{itemize}

All these methods, by construction, lead to a deadlock because the lack of information on the true emissivity prevents determining appropriate constraints to lead to the correct solution. The constraints claimed to be effective in leading to the true solution, in particular those discussed in Ref. \citenum{xing2016emissivity,sun2020constraints,
wang2021constraint,luo2022emissivity,
zhang2022multispectral,zhang2022ridge,zhu2022true,
gao2024method,yang2025data} are, in reality, just deluding ourselves. Again, since, in principle, no information about the true emissivity has been introduced that could help distinguishing it from all other \emph{permitted} solutions, it is pure logic to say that no kind of constraint will help to lead to it with certainty and systematically.

Refs.~\citenum{zhao2023approach} and~\citenum{zhang2024fast} are based on the construction of a matrix of radiance temperatures obtained by densely scanning the whole range of emissivity and the exploration for similarities in temperature values to find the (purportedly unique) solution of the temperature merging criterion in Equation~(\ref{eq:cost_T_i}). If was shown in Ref.~\citenum{krapez2025commentOE} that, depending on the matrix construction and search strategy, any \emph{permitted} solution could actually be found, thus demonstrating the absence of link with the true solution anyway. 

\subsection{Other MWT-NEI methods}
In this part, we provide the list of all other MWT-NEI methods not relying on the \emph{emissivity-way} approach. The diagnosis of temperature evaluation inefficiency made before applies to them as well. Indeed, since, according to the authors of each of these works, the criteria implemented in the optimization search they have developed have not been designed with reference to the properties of the "true" solution (which is the essence of the "NEI" perspective), they cannot help to find it accurately and confidently, except under trivial VPTR conditions. This strong deficiency will not be repeated systematically after citing each reference.

\subsubsection{Regression and nonlinear mapping}
Regression performed directly on radiance and nonlinear mapping from spectral intensity measurement to temperature distribution have been considered while applying: 
\begin{itemize}
\item Mixed Kernel Support Vector Regression (SVR) with Bayesian optimization of the SVR super parameters~\cite{zou2022multi},
\item Neural Networks~\cite{han2022cutting},
\item Ensemble Network structure comprising a base-learner (convolutional neural network (CNN), general regression neural network (GRNN), and BP) and a meta-learner (BP)~\cite{yang2024temperature},
\item algorithm that combines Transformer,  (Long Short-Term Memory (LSTM), and Support Vector Machine (SVM)~\cite{cui2025multispectral}.
\end{itemize}

\subsubsection{Generalized inverse matrix (GIM)}
After taking the logarithm of the radiance equations and applying Wien's approximation, we obtain a linear system in the unknown parameters; this linear system is underdetermined. The so-called \emph{generalized inverse matrix} (GIM) is a common mathematical tool for finding a particular solution to underdetermined linear systems. A few MWT-NEI methods use it :
\begin{itemize}
\item Combination of the Generalized
Inverse Matrix and the Exterior Penalty Function (GIM-EPF) applied on the EW cost function~\cite{liang2018generalized}
\item Generalized Inverse Matrix Normalization (GIM-NOR) algorithm~\cite{xing2020generalized}, which actually relies on empirical temperature determination, not optimization
\item Generalized Inverse Matrix followed by a Neural Network (GIM-RNN) trained on 243 emissivity models~\cite{xing2022generalizedOEng}
\item Generalized Inverse Matrix combined with a long short-term memory neural network (GIM-LSTM)~\cite{xing2022generalizedOE}
\end{itemize}
However, the solution provided by GIM is the one with the minimal norm (i.e. the minimal norm of the vector of the logarithm of emissivity). This means that some emissivity values are less than 1 but others are \emph{greater than 1}, which is impossible by definition of emissivity. The GIM solution is therefore inconsistent. For this reason, it was used as a starting point to search the "true" solution via a series of empirical and even graphical methods as recalled above. Let us mention that starting with the emissivity spectrum associated to the lowest permitted temperature, $T_L$ (see, e.g. the blue curves in Fig.~\ref{fig:Case_S_A_B_solutions}), would have been more physically justified. Anyway, whatever the starting point, in the absence of information on "true" emissivity, the search for it is nothing but blind and with unpredictable error, as with any other MWT-NEI method.

\subsubsection{Graphical methods}
Starting from the pyrometer signal versus wavelength data ($a_i$, $i=1,n$), which are essentially one-dimensional data, an image, i.e. two-dimensional data, can be built by computing for each cell the normalized quantity $(a_i-a_j)/(a_i+a_j)$, $i,j=1,n$. A series of work were published on the application of image recognition tools to extract the emissivity features and temperature from this 2D representation (in some instances, the Generalized Inverse Matrix was used to initiate the process):
\begin{itemize}
\item Generalized Inverse Matrix followed by a graphical deep learning neural network~\cite{zhang2023generalized,xing2023graphical,
zhang2024multispectralFusion} 
\item Generalized Inverse Matrix and improved multi-branch convolutional network model~\cite{zhang2025multispectral}
\item improved image classification algorithm integrating Squeeze-and-Excitation (SE) modules and convolutional neural network, CoAtNet~\cite{xing2025coatnet}
\end{itemize}
Changing from the 1D representation of the signal to the 2D representation defined above does not introduce any new information on the true emissivity. Since it is impossible to invert properly the temperature from the 1D representation when no prior knowledge of emissivity is available, there is no chance for a efficient inversion from the 2D representation.

\subsubsection{Tikhonov regularization}
Another avenue was explored to solve the underdetermined linear system, it is based on least squares and regularization.
Tikhonov regularization, also known as ridge regression, is most often implemented to solve ill-posed overdetermined problems, namely when the independent variables are highly correlated. It was applied to MWT in Ref.~\citenum{zhang2022ridge} and to multiwavelength thermography in Ref.~\citenum{sauer2019numerical}. This regularization method is effective for extracting the true solution (actually a biased approximation of it) amid other solutions also explaining the observed variables within a given error, taking advantage of the fact that the former has regularity characteristics that the others don't. This regularity is usually translated through the gradient or the Laplacian. 
The difficulty, in the context of MWT, is that all \emph{permitted} solutions have approximately the same regularity or smoothness (see Fig. \ref{fig:Case_S_A_B_solutions}, \ref{fig:Case_R_C_D_solutions}, \ref{fig:Case_M_2_3_solutions}, or \ref{fig:Case_M_1_2_3_solutions_and_T2_T3}-left). Implementing the Tikhonov regularization would favor one solution over another on the highly debatable basis of greater smoothness, but what guarantees that the extracted solution matches, or at least is closest to, the true solution sought? The algorithm extracts the \emph{permitted} solution that best satisfies the Tikhonov regularization criterion, that's all.

\subsubsection{MWT-NEI models based on local greybody assumption}

The MWT-NEI method developed in Ref.~\citenum{zhao2021multispectral} takes advantage that the measurement is performed at a high number of wavelengths and assumes that emissivity has slow variations. A moving spectral window is introduced and a temperature value is deduced for each window position based on the hypothesis that in each window the emissivity is \emph{constant} (locally grey hypothesis). The temperature values obtained by moving spectral window are then averaged to yield the estimated temperature. The method described in Ref.~\citenum{yan2025experimental} is essentially the same.

Now imagine that the 8 radiance data in Fig.~\ref{fig:Case_S_A_B_Rad} are interpolated to simulate a multiwavelength measurement with $n\gg8$ data. The methods in Ref.~\citenum{zhao2021multispectral} and Ref.~\citenum{yan2025experimental} would each provide a single temperature value. However, there is an infinity of \emph{permitted} temperature solutions, between $\hat{T}_L$ and $\hat{T}_U$, with the corresponding emissivity spectra obtained after filling the 8-point spectra in Fig.~\ref{fig:Case_S_A_B_solutions}. The greybody assumption is just a misleading hypothesis that denies the existence of so many potential exact solutions, among which the \emph{true} solution remains hidden.

\subsubsection{Multitemporal MWT}
\label{subsubsec:Multitemporal}
Multitemporal MWT where temperature evolves and emissivity is allowed to change with time, temperature, oxydation states or phase composition, is nothing else than an association of independent MWT problems. It is clear that at each time step the measured spectral radiance data give rise to a set of \emph{permitted} solutions encompassing temperature and spectral emissivity. The \emph{permitted} temperature values are in a range that may evolve with time: $[\hat{T}_L(t),\hat{T}_U(t)]$. The procedure consisting in picking, at each time, a temperature value in the corresponding  \emph{permitted} temperature range, together with the associated emissivity spectrum yields a sequential meta-solution that explains perfectly the observed multitemporal radiance data. Nevertheless, random picking gives rise to erratic temperature and emissivity variations (yet radiometrically plausible if we only consider the inferred radiance). Another option is picking in such a way that the temporal variations of emissivity and temperature are smooth, which is closer to most of expected physical phenomena. In this context, an inversion method where emissivity changes in a smooth fashion with time was proposed in Ref.~citenum{hagqvist2014emissivity}. The starting point is an initial solution assumed to be exact, i.e. either the initial compound emissivity is known (case a) or the initial temperature is known (case b). In case b, the new radiance set is used to calculate a new emissivity spectrum by taking the temperature at the previous step. Then, or for case a, the new radiance set is used to calculate a new temperature by taking the emissivity spectrum at the previous step; the same radiance set and the new temperature are used to infer an updated emissivity spectrum; the process is then iterated. What can be said against this method is that even though we start from an exact solution, the calculated time-varying meta-solution is just \emph{one} slowly varying solution among the huge amount of other slowly varying \emph{permitted} meta-solutions. Even though the calculated time-varying solution is "anchored" to reality at time $t=0$ (which, in itself, is already difficult to accomplish), nothing prevents it to diverge from the true emissivity and temperature as time progresses. This is simply due again to the non-uniqueness of the solutions at each individual time step. The diversity of meta-solutions is hidden from anyone applying the algorithm due to the deterministic nature of the explicit scheme used for the time progression of temperature and emissivity. As a consequence, the sources of uncertainty evaluated in Ref.~\citenum{hagqvist2014emissivity_sens} are highly underestimated because of overlooking the underdeterminacy of the MWT problem.

\subsection{Multiband thermometry (MBT)}
When the measurements are performed over very narrow spectral bands, the individual signals are obtained by multiplying the radiance in Eq.~(\ref{eq:Radiance}) by the bandwidth. When the bands are large, we then speak of multiband or multispectral thermometry (MBT, MST). The product of emissivity, blackbody radiance and detector spectral sensitivity must therefore be integrated over each wide band. As a result, each signal depends on intraband variations in emissivity, not only on the mean emissivity in this band. This is exacerbated by the fact that the spectral distribution of blackbody radiance in the band depends on temperature. As a result, additional difficulties arise compared to MWT, as will now be demonstrated.

MWT presents difficulties due to the multiplicity of solutions. For a given set of spectral radiance data there is an infinity of sets of emissivity spectra (and thus of temperatures) that provide the same measured data. At each wavelength $\lambda_i$, there is an infinite number of combinations of emissivity $\epsilon(\lambda_i)$ and blackbody radiance $B(\lambda_i,T)$, hence temperature $T$, that give the same measured signal $S_i$. This is the explanation for the deadlock when no information about emissivity is available. In the case of MBT, it is easy to imagine that the multiplicity of solutions is even higher. As a matter of fact, we have a secondary multiplicity of solutions originating from the fact that for any temperature value, there is an infinity of emissivity functions $\epsilon(\lambda)$ leading to the same value of the integral of the product of emissivity, blackbody radiance and spectral response, hence the same value of the spectral band signal. The fact that spectral bands can overlap introduces a continuity constraint on the emissivity function between consecutive bands, but the solution multiplicity persists. As a consequence, the absence of any information on the emissivity spectrum once again makes any attempt at accurate temperature assessment futile. This casts serious doubt on the ability of MBT-NEI methods, among which the tri-color methods~\cite{li2023color,yu2024theoretical,yu2024construction,
yu2024optimal,wei2024research}, to provide accurate temperature. A future blind test dedicated to MBT-NEI will dispel these doubts (if there were any left).

\subsection{About the proper and improper use of emissivity models}
\label{subsec:emissivity_models}
\subsubsection{Interpretation of Least Squares results}
\label{subsubsec:Interpretation}
There is a extensive literature on the use of emissivity models to solve the MWT problem (see e.g. the references cited in Refs.~\citenum{krapez2011measurements,araujo2017multi}. A possible way of eliminating underdeterminacy of the MWT problem is to introduce a relationship between the $n$ unknown spectral emissivity values, for example by means of an analytic parametric function $f_\textbf{P}(\lambda)$ describing emissivity as a function of wavelength, i.e.
\begin{equation}
\label{eq:function}
\epsilon(\lambda_{i})=f_\textbf{P}(\lambda_{i})\,,\;i=1,...,n
\end{equation}
where $\textbf{P}$ represents the vector of parameters, their number being between 1 and $n-1$ at most.
The one-parameter case corresponds to the well-known \emph{grey} model aimed for so-called greybody surfaces, i.e. with constant emissivity (equal values at the $n$ wavelengths of the measurement; notice that, in between, true emissivity can take different values). Other common analytical models are the linear model, the polynomial model and the exponential fonction of these (the latter makes that the logarithm of emissivity is expressed as a linear or polynomial function of wavelength). The grey-band model was also studied, namely a discrete model where emissivity takes a common value at two or more wavelengths, and such clustering can be met one or several times (see the discussion on this topic in Ref.~\citenum{krapez2011measurements}).

In the literature, no consensus is seen regarding the optimal choice for the model. Success is variable: depending on the material, the surface state, and the chosen model, the temperature error can be high or low.
Nevertheless, the general feeling is that the model should closely conform with the shape of the true emissivity spectrum and that it is preferable to limit the number of free parameters to no more than three, or even two.
As an example to avoid, we have the polynomial model of degree $n-2$ (so with $n-1$ free parameters). The degree is just low enough to eliminate the underdeterminacy, and, in the same time, high enough to approximate a wide variety of real spectra defined on $n$ values, all the more so as $n$ is large. Nevertheless, this polynomial model turns out to be a very bad candidate, especially for large values of $n$. Coates showed in Refs.~\citenum{coates1981multi,coates1988least} that the temperature errors can then take unpredictably high values.

A new perspective that considerably helps to interpret and compare the results obtained with different emissivity models and thus to implement them successfully was described in Ref.~\citenum{krapez2011measurements,
krapez2019measurements}. We recall the main aspects and focus on those that help to understand the failure of MWT-NEI methods involving emissivity models.

The determination of the optimal values of the model free parameters and, simultaneously, that of temperature, is usually performed by minimizing the cost function (least-squares method - LS):
\begin{equation}
\label{eq:Xi2}
\min_{\textbf{P},T} \sum_{i=1}^{n}\left[L_i-f_\textbf{P}(\lambda_{i}) B(\lambda_{i},T) \right]^2.
\end{equation}
Let us factorize by $B(\lambda_{i},T)^2$ and transform the expression to change from a joint minimization on temperature and the parameters of the model $f_\textbf{P}(\lambda)$ to a sequential minimization. For a reason that will soon become clear, we replace the symbol $T$ by $\hat{T}$. Eq.~(\ref{eq:Xi2}) is thus changed into:
\begin{equation}
\label{eq:Xi2_a}
\min_{\hat{T}} \left\lbrace  \min_{\textbf{P}} \sum_{i=1}^{n}B(\lambda_{i},\hat{T})^2\left[\frac{L_i}{B(\lambda_{i},\hat{T})}
-f_\textbf{P}(\lambda_{i})\right]^2\right\rbrace.
\end{equation}
The result of this minimization is of course the same as in Eq.~(\ref{eq:Xi2}). Now let us focus on the internal minimization (i.e. the one in braces). It means that for a trial temperature $\hat{T}$ the minimization is performed just on $\textbf{P}$:
\begin{equation}
\label{eq:Xi2_b}
\min_{\textbf{P}} \sum_{i=1}^{n}B(\lambda_{i},\hat{T})^2\left[\frac{L_i}{B(\lambda_{i},\hat{T})}
-f_\textbf{P}(\lambda_{i})\right]^2.
\end{equation}
We can recognize in the first term in square brackets the trial emissivity spectrum related to $\hat{T}$ (see Eq.~(\ref{eq:permitted_emissivity})). Eq.~(\ref{eq:Xi2_b}) becomes:
\begin{equation}
\label{eq:Xi2_c}
\min_{\textbf{P}} \sum_{i=1}^{n}B(\lambda_{i},\hat{T})^2\left[\hat{\epsilon}_{\hat{T},i}
-f_\textbf{P}(\lambda_{i})\right]^2.
\end{equation}
So, the internal minimization corresponds to a weighted curve fitting of the trial spectrum $\hat{\epsilon}_{\hat{T},i}$ by the parametric function $f_\textbf{P}(\lambda)$. The weight is $B(\lambda_{i},\hat{T})^2$, the square of blackbody radiance spectrum at temperature $\hat{T}$; it can be considered that it plays the role of the reciprocal of the variance in classical weighted least squares (WLS).

The global minimization in Eq.~(\ref{eq:Xi2_a}) is achieved by nesting the internal minimization with a minimization according to $\hat{T}$:
\begin{equation}
\label{eq:Xi2_3}
\min_{\hat{T}} \left\lbrace  \min_{\textbf{P}} \sum_{i=1}^{n}B(\lambda_{i},\hat{T})^2\left[\hat{\epsilon}_{\hat{T},i}
-f_\textbf{P}(\lambda_{i})\right]^2\right\rbrace.
\end{equation}

On the other hand, when the radiance equations are first transformed by taking the logarithm, LS becomes:
\begin{equation}
\label{eq:Xi2_lin}
\min_{\hat{T}} \left\lbrace  \min_{\textbf{P}} \sum_{i=1}^{n}\left[\ln(\hat{\epsilon}_{\hat{T},i})
-f_\textbf{P}(\lambda_{i})\right]^2\right\rbrace.
\end{equation}
where $f_\textbf{P}(\lambda)$ is now a model function describing the logarithm of emissivity. In this case, weighting is absent from the LS expression.

We now have the elements to clarify what is actually achieved with radiance LS and dispel erroneous thoughts. 
\begin{itemize}
\item what the LS method does: ultimately, the minimization at stake, whether on radiance or its logarithm, comes down to finding the spectrum $\hat{\epsilon}_{\hat{T},i}$, resp. its logarithm, and the corresponding temperature $\hat{T}$, for which the parametric function $f_\textbf{P}(\lambda)$ gives the lowest minimum residual norm. In other words, it yields the \emph{permitted} solution that can best match with the chosen model function, either constant, linear, polynomial or anything else. 
\item what the LS method does not: a misinterpretation of LS is to expect that in some way the \emph{true} emissivity spectrum is first fitted with the chosen parametric function $f_\textbf{P}(\lambda)$ and then, based on this spectrum approximation, the temperature is calculated so that the inferred radiance spectrum matches at best with the observed one. Of course, this desirable operation is only a dream since the \emph{true} emissivity spectrum is unknown to allow the first operation.
\end{itemize}

An imaginary process can be put forward to explain the LS results. It is as if the parametric function $f_\textbf{P}(\lambda)$ were used to fit each \emph{permitted} spectrum, one by one, and only the one for which the residual norm is minimal is retained, together with the corresponding temperature. In this virtual process, and assuming error-less data, the \emph{true} spectrum would also happen to be fitted, but anonymously, and if the residual norm is not minimal, it is simply bypassed and forgotten.
In any case, LS does not consist of a fit of the \emph{true} spectrum alone by the model $f_\textbf{P}(\lambda)$, and for good reason: the \emph{true} spectrum is unknown to us at this stage.

The essential point to remember is that in the present MWT application, LS is a \emph{shape-finder}, not a \emph{true-solution finder}. 

\subsubsection{Recommended strategy for finding the true solution}
\label{subsubsec:Recommended}

As a consequence of the interpretation given before and based on the relationship between all \emph{permitted} spectra described in Eq.~(\ref{eq:permitted_emissivity_1_and_2_Wien}),
only when the \emph{true} spectrum $\epsilon_{true, i}$ better matches with the target shape (i.e. the one generated by the parametric model, i.e. flat or linear or parabolic, etc...), than do all other \emph{permitted} spectra $\epsilon_{true, i}\;\exp(\alpha/\lambda_i),\;i=1,...,n$, where $\alpha$ is any non-zero constant, does LS provide the true temperature with high accuracy. This is in the hypothetical case of error-free measurements. Since random errors are always present, both in the radiance measurements and the independent emissivity measurements, the notion of better or worse fit and the final uncertainty must be judged with reference to the noise level.

We can build up on this observation to define a procedure for selecting the parametric model $f_\textbf{P}(\lambda)$ likely to yield optimal results:
\begin{itemize}
\item have access to experimental data describing the true emissivity profile, not necessarily at the same wavelengths as those of the MWT experiment, but encompassing its range, say $\epsilon_{exp, j},\;j=1,...,J$
\item find a parametric function $f_\textbf{P}(\lambda)$  such that the goodness of fit is good for $\epsilon_{exp, j},\;j=1,...,J$, and better than for any other set
\begin{equation}
\label{eq:criterion_1}
\epsilon_{exp, j}\;\exp(\alpha/\lambda_j)\,,\;j=1,...,J,
\end{equation}
where $\alpha$ is a positive or negative constant (application of the Wien's approximation)
\item apply the selected model $f_\textbf{P}(\lambda)$ for the LS problem related to the radiance values of the MWT problem, i.e. $L_i,\;i=1,...,n$ and obtain an estimate of the true temperature $T_1$.
\item if needed, a refinement can be made by putting aside the Wien's approximation. The model selection is updated by considering Eq.~(\ref{eq:permitted_emissivity_1_and_2}) instead of  Eq.~(\ref{eq:permitted_emissivity_1_and_2_Wien}), now that a temperature estimate is known. The selection criterion becomes: find a parametric function $f_\textbf{P}(\lambda)$  such that the goodness of fit is of high quality for $\epsilon_{exp, j},\;j=1,...,J$, and better than for any other set 
\begin{equation}
\label{eq:criterion_2}
\epsilon_{exp, j}\;\frac
{\exp\left(\frac{C_2}{\lambda_{j}(T_1+\Delta T)}-1\right)}
{\exp\left(\frac{C_2}{\lambda_{j}T_1}-1\right)}
\,,\;j=1,...,J,
\end{equation}
where $\Delta T$ is any positive or negative temperature offset.
\item if another model has thus been selected, apply it to the LS problem to obtain a better estimate of the true temperature.
\end{itemize}

Notice that if it happened that the chosen model fitted better a particular set $\epsilon_{exp, j}\;\exp(\alpha/\lambda_j)\,,\;j=1,...,J$, than the expected spectrum, $\epsilon_{exp, j}\,,\;j=1,...,J$, (the model was thus badly chosen), instead of providing the true temperature $T_{true}$, LS would provide the temperature $T_{\alpha}$, which is related to the true one by:
\begin{equation}
\frac{1}{T_{\alpha}}=\frac{1}{T_{true}}+\frac{\alpha}{C_2}.
\end{equation}
Of course, the procedure described above can only be implemented if preliminary efforts have been made to get information on true emissivity (no pain, no gain!). This information is necessary to identify a parametric function $f_\textbf{P}(\lambda)$ that fits well the \emph{true} spectrum $\epsilon_{true, i}$, and in the same time fits all other \emph{permitted} spectra less well.

In the literature, there is no well established criterion for model selection. Vague recommendations are sometimes given, such as finding a model that allows fitting the \emph{true} emissivity spectrum, but at the same time has a small number of parameters~\cite{rusin2018determination}. This is in line with the criteria expressed above with Eq.~(\ref{eq:criterion_1}) and Eq.~(\ref{eq:criterion_2}). Indeed, by keeping the number of parameters low, unwanted \emph{permitted} solutions are more easily rejected, because they show a higher residual norm. The new criteria in  Eq.~(\ref{eq:criterion_1}) and Eq.~(\ref{eq:criterion_2}) do, however, express the objective more explicitly. In practice, the residual norm should be minimal for the \emph{true} emissivity spectrum, and increase rapidly as we move away from it in the solution space.

The use of LS in MST presents many traps (illustrations with numerical examples on the benefits and pitfalls of using emissivity models can be found in Ref.~\citenum{krapez2011measurements,
krapez2019measurements}). In particular, careful selection of the emissivity model is of prime importance and overlooking it can have serious consequences: not applying the criteria above at all exposes us to get accurate results only by chance or in trivial conditions of VPTR.

\subsubsection{Alternate strategies and consequences}
\label{subsubsec:Alternate}
The optimal selection of the parametric function $f_\textbf{P}(\lambda)$ as described before is quite demanding in terms of prior knowledge of emissivity. Good quality experimental data are required. In their absence, other strategies have to be devised, which can have adverse consequences on temperature accuracy. For example, we may be faced with the following dilemma when selecting a model:
\begin{itemize}
\item if the parametric function has too few free parameters, there is a high risk that the internal minimization of the LS problem will produce a higher residual norm for the true solution than for other \emph{permitted} solutions. LS will thus converge on one of the latter, and the temperature obtained may be very far from the true temperature. Small variations in the radiance signal, as due to measurement noise, will not change substantially the result
\item if the parametric function has too many free parameters (as for example a polynomial of high degree), the internal minimization of the LS problem will produce low residual norm for a large range of \emph{permitted} solutions (this is because the model was chosen to be very versatile, likely too versatile), among them probably the true solution (if the model was not too badly chosen with respect to it). The LS method will then converge on one among all of these, and the obtained temperature will be unpredictably far or close to the true temperature. In addition, small variations in the radiance signal, as due to measurement noise, will bring dramatic and unpredictable changes to the result. 
\end{itemize}

The danger in using too versatile models (e.g. polynomials of high degree) was illustrated by Coates in Ref.~\citenum{coates1981multi,coates1988least}. In Ref.~\citenum{wang2018new} the degree of the polynomial model was 7. The experimental emissivity data were successfully fitted with a polynomial of degree 7, however, performing LS with a polynomial of such a high degree exposes finding an unacceptably high uncertainty for the calculated temperature. The same criticism applies to the use of multi-segment linear models~\cite{zhang2022data}, because again of the high number of free parameters, which makes that a large number of \emph{permitted} solutions meet the criterion of a good fit with the parametric function, making it difficult to extract the true solution from among them.

When nothing is known about emissivity, the first reflex is to apply the grey model, hoping that this simplifying assumption is not too restrictive and does not have too many consequences. In fact, in the MWT literature, we can find innumerable papers adopting the greybody assumption arguing in particular that because of the short wavelength range, emissivity should not vary too much.
On the other side, when we have to approximate a function showing variations, it is usual to start with the first-order approximation, i.e. the linear approximation. Often, this proves to be sufficient. In the present context of LS applied to MWT, this way of thinking is misleading as will be shown next.

Actually, whatever the emissivity model considered for LS (constant, linear, ...), our recommendation is first to draw the bundle of \emph{permitted} solutions and check if one of them can be properly fitted with this model. If this is not the case, change the model or increase the number of free parameters. Otherwise, we cannot trust the temperature identified by LS. As an example, the grey model and the linear model are both unsuitable to solve the problem plotted in Fig.~\ref{fig:Case_S_A_B_solutions}-left. 

However, what is important to remember is that, in the opposite situation, namely if a \emph{permitted} spectrum proves to be well fitted with the chosen model, this does not mean that the temperature given by LS is the true one. Let us explain it in detail with an example.

Consider the blind test for scenario S and material B, but only for the limited spectral range [1.4 µm, 2.03 µm], i.e. on the left of the discontinuity shown in the \emph{permitted} spectra in Fig.~\ref{fig:Case_S_A_B_solutions}-right. Let us now assume that, for whatever reason, the linear emissivity model has been chosen for the LS method. LS gives the following result: 1050 K. Are we ready to accept this value as the true one ? What does the LS result actually mean ? It simply means that the \emph{permitted} spectrum closest to a straight line is that corresponding to 1050 K. That's all. This result certainly does not tell us that the \emph{true} temperature is 1050 K. This, even if the linear fit with the \emph{permitted} spectrum corresponding to 1050 K is here perfect.

The discussion in § \ref{sec:Blind_test_S} highlighted the fact that the \emph{true} temperature could be any value between $\hat{T}_L$=997.9 K and $\hat{T}_U$=1285.2 K. 
Performing LS with a linear model does not change anything: we are still unable to tell what the \emph{true} temperature is. Only if there is experimental and independent evidence that the \emph{true} spectrum $\epsilon_{true, i}$ fits well with a linear function, better than any other spectrum $\epsilon_{true, i}\exp(\alpha/\lambda_i)$, can we state that \emph{true} temperature is 1050 K (in practice, there is a temperature uncertainty; it can be estimated from the residual norm and the experimental errors). Otherwise, we have no way of telling where it falls within the range [$\hat{T}_L$, $\hat{T}_L$]. Of course, this example can be transposed to any other parametric model.

In conclusion to this illustration, a vague idea of the shape of the true emissivity spectrum is not enough and taking a model for granted without prior experimental confirmation exposes us to large errors. Above all, succeeding in reducing the LS residual norm to a very low level is in no way a guarantee of success in the search for true temperature. This is just a first step that must be completed with information on true emissivity.

It is easy to deduce from all the above that \emph{blind} application of a parametric model $f_\textbf{P}(\lambda)$ must be strictly avoided, regardless of whether the number of parameters is small or not. Proscribing the approach consisting in \emph{blind} picking among a series of models and retaining one of them on the sole basis of low LS residuals proceeds from the same analysis. A few examples of such misuse of emissivity models will be described in the next paragraph.

In the field of MWT-LS, it is important to make the difference between emissivity models applied \emph{blindly} (EMAB) and emissivity models applied \emph{knowingly} (EMAK).
By principle, MWT-NEI methods fall in the EMAB category since nothing is known on emissivity, and therefore on its shape. It is then impossible to choose a parametric model that is perfectly adaptable to the \emph{true} emissivity spectrum.

\subsubsection{Emissivity models applied blindly - EMAB}
\label{subsubsec:EMAB}
\paragraph{Grey model applied blindly}
\label{paragraph:Grey-EMAB}
 
The simplest model is the grey model which assumes that emissivity is the same at all considered wavelengths. There are many examples of its use, mainly in bi-color thermometry (BCT) (see e.g. Ref.~\citenum{yamada2003thermal,ketui2016single,hooper2018melt,
dicarolo2020standard,vecchiato2020melt,
vallabh2021single,schwarzkopf2023two,
myers2023high,glasziou2024measurement,alam2025enhancing,
yang2025temperature,yu2025monitoring,dicarolo2025spectral,
bayat2025enhanced}, but also in WMT (see e.g. Ref.~\citenum{ni2024flame,hameete2024particle
,dong2024multi,cao2025thermometry,pei2025research,
bykov2025four}). Typical errors associated with its use will now be described.

Firstly, the grey hypothesis is often adopted by default, due to poor knowledge of the radiative properties of the surface~\cite{huang2019effect,glasziou2024measurement
,pei2025research,bykov2025four}. As we saw earlier, this attitude should be avoided. Secondly, an argument often put forward to justify the application of the grey hypothesis, particularly in the context of BCT, is as follows: " by choosing wavelengths that are close, it can be assumed that the emissivities are the same"~\cite{hooper2018melt}, or "although no real surface is truly gray, it often happens that the emissivity is relatively constant along a narrower wavelength range of interest"~\cite{araujo2014monte,araujo2015monte,araujo2016analysis,
araujo2016dual}, or "the bands are typically selected very close to each other to reduce the approximation error (local gray-body assumption)"~\cite{fagnani2024line}, or "although the
gray body assumption is not valid for metallic surfaces, through
the choice of appropriate two wavelengths, very close to each
other, the spectral dependence of emissivity can be neglected"~\cite{musto2016error}, which, correlatively, prompts to say "when increasing band separation the gray surface assumption becomes less likely"~\cite{araujo2014monte,araujo2016analysis,araujo2016dual}.
These arguments are not correct (sorry about that, an equally erroneous argument was expressed in Ref.~\citenum{pierre2022simultaneous} where it was written that the “greybody” approximation, to be satisfactory, requires that the two wavelengths not be too far apart).
Of course, when $\lambda_1$ comes close to $\lambda_2$, then $\epsilon(\lambda_1)$ comes close to $\epsilon(\lambda_2)$ as well. This is simply the expression of function continuity, which can be considered as valid for solid materials. By dividing the radiance equation Eq.~(\ref{eq:Radiance}) expressed at each of the two wavelengths, the result involves the ratio of the two emissivities and the observation made just before makes many authors replace it by 1 (e.g. Ref.~\citenum{yamada2003thermal,hooper2018melt,
dicarolo2020standard,vecchiato2020melt,
vallabh2021single,schwarzkopf2023two,
glasziou2024measurement,alam2025enhancing,
yang2025temperature,yu2025monitoring,dicarolo2025spectral,
bayat2025enhanced,pei2025research}). The equation is then free of emissivity values, the temperature can therefore be deduced easily, as if the surface was a blackbody, by the way. This approach is at the heart of bicolor ratio thermometry.
However, that $\lambda_1$ comes close to $\lambda_2$ (and $\epsilon_1$ to $\epsilon_2$) does not justify that we can apply the grey hypothesis and consider the inferred temperature as the true one. Indeed, replacing the emissivity ratio by 1 is a trick, actually a trap, that makes us adopt the \emph{permitted} solution that shows $\hat{\epsilon}_1=\hat{\epsilon}_2$ and forget all others.

As an illustration, consider any pair of two close wavelengths $\lambda_i$ and $\lambda_{i+1}$ in Fig.~\ref{fig:Case_S_A_B_solutions}-left or right. First of all, it should be noted that when only two wavelengths are taken into account, instead of all eight, the range $[\hat{T}_L ,\,\hat{T}_U]$ of \emph{permitted} temperature is often larger. Next, for some pairs of wavelengths, we can find a temperature $\hat{T}$ for which $\hat{\epsilon}_{i}=\hat{\epsilon}_{i+1}$. This temperature would be that obtained by applying the grey hypothesis. However, this would make us forget that the \emph{true} temperature can be any temperature between $\hat{T}_L$ and $\hat{T}_U$. Furthermore, and this is the key point, bringing the two wavelengths closer would not change anything to the previous sentence (the only possible change is a slight modification in the values of $\hat{T}_L$ and $\hat{T}_U$). The wavelength-proximity criterion to justify the grey hypothesis is therefore misleading. Bringing the two wavelengths close increases the probability that the two emissivities are close but \emph{this does not allow saying that they are equal} and that the grey-solution temperature is the true one.
In reality, the only way to justify the grey hypothesis is to \emph{verify experimentally} that the two emissivities are equal. Without this verification, any application of the grey hypothesis is therefore done \emph{blindly}, hence at your own risk. It is possible that the result is nevertheless correct, but this is due to chance.

In Ref.~\citenum{ni2024flame,hameete2024particle
,cao2025thermometry} the grey hypothesis was applied  to process MWT measurements. The Least-Squares method was applied and, in Ref.~\citenum{cao2025thermometry, hameete2024particle}, the inferred temperature was accepted without, apparently, checking the related emissivity spectrum against the grey hypothesis. Actually, the simple fact that the norm of LS residuals is small is in no way a validation criterion for the model used, as explained in §~\ref{subsubsec:Interpretation}. In Ref.~\citenum{ni2024flame}, on the basis on the temperature identified, the deduced emissivity spectrum (according to Eq.\~(\ref{eq:permitted_emissivity})), was verified to be nearly constant, at least over part of the considered spectral band. Based on this observation, the grey hypothesis was justified a posteriori. This is not correct. In fact, the correct interpretation, as already mentioned, is that the temperature thus identified is that of the particular \emph{permitted} solution showing an almost constant emissivity on this part of the spectrum, nothing else. Claiming that this is the \emph{true} temperature is a gamble we wouldn't recommend to do.

In Ref.~\citenum{dong2024multi} the grey hypothesis was applied for each two successive wavelengths leading to $n-1$ radiance ratios deprived from emissivity, hence leaving only blackbody radiance ratios. These were then averaged and a temperature value was derived thereupon. Theoretically, the true temperature is obtained in the pure grey case, i.e. a constant emissivity for all considered wavelengths. The error due to the departure from the grey hypothesis for one or more pairs of wavelengths is unpredictable (we can't be satisfied with a validation carried out only on blackbodies~\cite{dong2024multi}).

Notice that these misuses are not due to the grey model itself, the methodologies described are wrong regardless of the model used (e.g. Ref.~\citenum{fu2011fast,fu2014temperature} also use the wavelength-proximity criterion to justify the use of a linear model; this is not correct, as it was for the grey model). For the methodology to lead to the true temperature, or close to it, it is necessary (but not sufficient) that the true emissivity spectrum could be well fitted with the chosen model. We have to admit that this prerequisite is not always easy to meet.

\paragraph{Other models applied blindly}
\label{paragraph:Other-EMAB}

MWT-NEI methods pretend not requiring a priori information on emissivity. This is of course in contradiction with the constraints imposed for an efficient use of emissivity models with the LS method. As we saw in §~\ref{subsubsec:Interpretation} and §~\ref{subsubsec:Recommended}, the recommended approach consists in choosing a model that can faithfully reproduce the shape of the \emph{true} emissivity spectrum (this is a necessary but not sufficient constraint). Therefore, in the absence of information on \emph{true} emissivity, how can it be possible to achieve the required capacity for faithful reproduction, except by chance? Yet, this highly improbable target was purportedly achieved in many papers~ \cite{chen2020multi,cai2024experimental,
wang2024multiReconstructed,yang2023development,
yang2025multispectral}.

In Ref. \citenum{chen2020multi}, first, the model was selected by a Backpropagation (BP) neural network from a series of seven models (linear, quadratic, sinusoidal, exponential, ...), then Least Squares was carried out by an Improved Non-dominant Sorting Genetic Algorithm (INGSA). By letting the algorithm select the emissivity model among a series of analytic models, we let it lead to the \emph{permitted} solution showing the smallest fit residual norm, all models combined. What allows us to say that the temperature obtained corresponds to the \emph{true} one, or at least is close to it ? Nothing. The elements developed so far in this paper allow us to say that there can be no link between the temperature obtained by this MWT-NEI method and the true temperature. If they turned out to be close to each other, it would be purely due to chance, to a situation of VPTR, or the application of unpublished constraints.

Ref.~\citenum{coffman2025spectral,yang2023development} used a linear model for resp. emissivity and the logarithm of emissivity. In both cases, the authors argued that the model they have chosen is common for the materials concerned, but without verifying that the true emissivity actually follows the chosen law. This is a common methodological error.

Ref.~\citenum{cai2024experimental,yang2025multispectral} used a polynomial to model emissivity but the choice of the degree of the polynomial was not even discussed (in Ref.~\citenum{guan2025tomographic} not even the type of analytical model was mentioned). Remember that Coates warned against using high-degree polynomials \cite{coates1981multi,coates1988least}.
Ref.~\citenum{zhang2025improved} is another example of a blind application of emissivity models: three to six channels of a multispectral camera were used to get radiance images of a C/SiC specimens submitted to an
oxy-propane flame ablation system, and a polynomial model of order $n-3$ was systematically used, which means, the maximum possible order minus one. Different interpretations were given for the consistency or inconsistency of the results, but what matters is the respective quality of the fit by the polynomials under consideration of the (unknown) emissivity spectrum of C/SiC compared to all other \emph{permitted} spectra.

A polynomial model was also used in Ref.~\citenum{mironov2024experimental}; the degree could be up to the highest possible value minus one, namely $n-3$, where $n$ is the number of wavelengths. Actually, although not mentioned in Ref.~\citenum{mironov2024experimental}, in practice, the polynomial degree was one~\cite{mironov2025private}. A moving window was used and the minimal LS residual norm thus obtained was claimed to yield the true temperature. However, given the blind application of the polynomial model (whatever the degree), this correspondence can only be fortuitous in the general case. Hopefully, the result was close to the true temperature, which is most probably due to the fact that the emissivity was very high (about 0.95) in the considered spectral band.
Similarly, a moving triplet of wavelengths was used  together with a linear model applied blindly for each triplet in Ref.~\citenum{fu2011fast,fu2014temperature}, and without any other justification than the erroneous one of the proximity of the three wavelengths, disregarding the existence of multiple solutions to the MWT problem related to each triplet.

In Ref.~\citenum{zheng2019measurement} the 2D distribution of temperature in a candle flame was obtained with a hyperspectral camera with up to 128 spectral bands by using a polynomial model of unknown degree. Later, the authors considered together the grey model, the Hottel and Broughton soot emissivity model, the Rayleigh model and a polynomial model (again of unknown degree)~\cite{zheng2021measurement} to process the signal from a spectrometer aiming at a biomass candle flame. 
In Ref.~\citenum{zheng2022measurement,zheng2022situ} the Hottel and Broughton model was considered alone for measuring again flame temperature and emissivity. In all these cases, the accuracy of the results depends strictly on the adequacy of the chosen model and the true emissivity spectrum of the flame, which was speculated, not demonstrated.

In the same spirit of avoiding blind application of models, the strategy applied in Ref.~\citenum{girard2014multiwavelength,
battuello2016characterisation,liu2016measurement,rusin2018determination} of sequentially testing polynomials of increasing order, is not recommended. Ref.~\citenum{liu2016measurement} used it to measure the distributions of temperature and wavelength-dependent emissivity of a laminar diffusion flame, whereas Ref.~\citenum{rusin2018determination} used it to measure the temperature of a tungsten sample.
In Ref.~\citenum{rusin2018determination}, for each polynomial order, the curve of the norm of residuals vs. $\hat{T}$ was compared with the estimated norm of measurement errors $\delta_{exp}$. If the curve was well above $\delta_{exp}$, the model was deemed inadequate, and the order of the polynomial was increased. For some value of the polynomial order, the minimum of the curve was just below $\delta_{exp}$; in this case, the range of temperature values $\hat{T}$ for which the residual norm is less than $\delta_{exp}$ was claimed by the authors to provide an estimate of the true temperature. Further increasing the polynomial order would just widen this temperature range (and make the minimum wander erratically according to Ref.~\citenum{coates1981multi,coates1988least}); the exploration should then stop. 
Our comment is that in doing so, we simply identify the situation (i.e. the polynomial order) where the range of \emph{permitted} spectra showing a good fit with the tested model is the smallest possible.
Since no information on the \emph{true} solution has been used, the result can have no deterministic connection with it. In Ref.~\citenum{rusin2018determination}, the validation was performed on simulated radiance data based on an experimental emissivity spectrum of tungsten taken from the literature. It turns out that a polynomial of degree 2 fits better the \emph{experimental} spectrum  then any other \emph{permitted} spectrum; so the true temperature was recovered with very little error. One chance. It seems peremptory to assert that such a contingency could occur with any other material and thus allow the method to be applied blindly.

Another simple rule was expressed in Ref~\citenum{girard2014multiwavelength,
battuello2016characterisation}, which consists in assuming as a reliable result when polynomials of two successive orders, for example 2nd- and 3rd-order, give
similar temperatures, namely within a specified threshold
value. This strategy works when the true emissivity spectrum closely fits with a polynomial function of low order. We believe that if this is not the case, the error can reach unpredictably high levels for the reasons mentioned before.
In Ref.~\citenum{liu2016measurement}, when increasing the polynomial order, the cost function decreased until reaching a plateau and the estimated temperature showed convergence. This last result is in contradiction with the findings in Ref.~\citenum{coates1981multi,coates1988least} where when increasing the polynomial order, the calculated temperature showed erratic variations of exponentially increasing amplitude.

Ref.~\citenum{picon2021probabilistic,
ennass2023temperature} used a continuous piece-wise linear model based on P1-hat functions with, resp. 10 and 3 free parameters. The validation in Ref.~\citenum{picon2021probabilistic} was performed on ceramic samples and thus benefited from the fact that at Christiansen wavelength, emissivity is very close to 1, which most probably gave rise to VPTR conditions. We expect that in other situations where emissivity is not so high, small amount of noise makes the optimization process wander erratically from one \emph{permitted} solution to another one, over a large temperature range, especially in the case of a large number of linear pieces in the model. Again, regardless of the number of linear parts, the fact that the method outcome comes close to the true solution is pure contingency, which can be alleviated only if it was established that the true emissivity spectrum, can be approximated, better than all other \emph{permitted} spectra, with a piece-wise linear function, a task difficult to fulfill.

In Ref.~\citenum{wang2024multiReconstructed} the spectrum associated to a temperature $\hat{T}_{mid}$ taken between $\hat{T_L}$ and $\hat{T_U}$ is first fitted with the sparsest Laurent polynomial of degree $n$ at most. This sparse Laurent polynomial (with less than $n-1$ terms) was then used as an emissivity model in a classical least square optimization. It can be expected that the result is in close relation with the temperature $\hat{T}_{mid}$ chosen at the beginning, but it has no obvious connection with the \emph{true} temperature. How could it be otherwise without any indication of the shape of the \emph{true} emissivity spectrum?

Truncated singular value decomposition (TSVD) is classically used to solve ill-conditioned problems where, for example, the parameters present high correlations. In Ref. \cite{zhang2025quarterframe, zhang2025multispectrum}, after linearizing the equations, introducing a linear emissivity model, and noticing that the inferred linear system (4 wavelengths, hence 4 equations and 3 unknowns) showed a still high condition number, it was decided to apply TSVD. This of course alleviated the ill-condition of the problem but why, in the present case of overdetermination, would TSVD necessarily lead to the true temperature preferably to any other \emph{permitted} solution? TSVD leads to a solution that satisfies a regularity criterion defined by the minimum singular value retained, but it is not possible to rely on such a criterion to extract the true temperature for sure from among all the other \emph{permitted} temperatures.

LS with a polynomial model of highest possible degree, namely $n-2$, was once shown to be a bad choice when $n$ is large~\cite{coates1981multi,coates1988least}. Ridge regression, which involves the norm of the free parameters of the polynomial, was therefore added to regularize the problem~\cite{araujo2020surface}. The otherwise observed overfitting was eliminated by implementing the regularization. However, once again, why should the recovered temperature be close to the \emph{true} temperature? On what basis can we expect the true (yet unknown) emissivity spectrum to comply better than all other \emph{permitted} spectra with the ridge regression criterion?

Multitemporal MWT was already addressed in §\ref{subsubsec:Multitemporal} where the inversion was performed without the help of an emissivity model. An alternative is to introduce an emissivity model whose parameters are allowed to change at each new time step. As an example, in Ref.~\citenum{lamien2024sequential}, a linear emissivity model was chosen and the temperature evolution was estimated through the solution of a state estimation problem within the Bayesian framework with two algorithms of the Particle Filter method namely Sampling Importance Resampling and Auxiliary Sampling Importance Resampling. Notice that the maximum likelihood estimation could be obtained at each time step by LS. According to the above, the result would correspond to the \emph{permitted} solution whose spectrum is closest to linearity, time after time. The question is whether the tested material really has a linear emissivity spectrum, otherwise the solution is biased. With the Particle Filter instead, another sequential solution is obtained, approaching, at each time, another of the infinite number of solutions. Since the internal constraints build upon the linear hypothesis, with no consideration of the true emissivity spectrum, it is reasonable to fear that, in general, the result is highly uncertain.

\subsubsection{Emissivity models applied knowingly - EMAK}
\label{subsubsec:EMAK}
The success with LS applied to MWT depends on the adequacy of the emissivity model with the \emph{true} emissivity spectrum. This requires a preliminary effort for the characterization of the material or at least for an investigation on the probable shape of the emissivity spectrum. A few papers report on these preliminary operations, which is a prerequisite for selecting an emissivity model capable of giving the actual temperature with accuracy.  
see e.g. Refs.~\citenum{qu2020temperature,uman2006fiber
,neupane2022development}.

\subsection{MWT-NEI successes: either chance or hidden constraints}
\label{subsubsec:MWT-NEI_successes}
We are not trying to hide the fact that, \emph{on occasion}, a MWT-NEI method may provide an accurate result for a particular material, and under particular conditions. Good for the user, but we would be deluding ourselves if we thought this success could be generalized to other materials, situations or temperature levels.
Surprisingly, however, the literature on these methods is full of inversion results with high accuracy, whether applied to simulated or experimental data. This comes in conflict with the results of the blind tests exposed in this benchmarking work. Actually, the present results would support the hypothesis of information bias outlined before.

Hidden constraints were applied in Ref.  \citenum{zhang2024fast} to force the MWT-NEI searching process to converge on the true solution, see Ref. \citenum{krapez2025commentOE}. The constraint was to limit the range $[min(\hat{\epsilon}_{i}), max(\hat{\epsilon}_{i})]$ of the trial emissivity search in Equation~(\ref{eq:T_i}) to that corresponding to the true emissivity spectrum $[min(\epsilon_{i}), max(\epsilon_{i})]$ that was used to generate the synthetic radiance data (provoked VPTR situation). In real life, however, the values $[min(\epsilon_{i})$ and $ max(\epsilon_{i})]$ are not easy to know. Furthermore, if they were available, then, a simple glance to the plot of \emph{permitted} solutions as, e.g. in Fig. \ref{fig:Case_S_A_B_solutions}, would immediately give the true temperature; there is no need of a complicated \emph{emissivity-way} optimization method.

A preliminary adaptation of the trial emissivity range in accordance to the real bounds was also applied in, e.g. Ref.~\citenum{xing2017directly,liang2018generalized,luo2022emissivity,
zhang2023optimizing}. A similar operation was carried out in Ref.~\citenum{torres2023radiometric} where the emissivity bounds were established in accordance with those of the material tested but in ideal conditions. The authors recognized that "defining this range is of upmost importance, since the smallest variations in these parameters have big variations in the resulting temperature estimations"~\cite{torres2023radiometric}. 

Ultimately, these MWT-NEI methods, that rely on a close matching of the emissivity search range with that of the emissivity of the material concerned, cannot be considered as MWT-NEI methods, but as MWT-WEI methods (i.e. \emph{with} emissivity information). Furthermore, this ironically shows that MWT-NEI methods cannot do without information on emissivity, hence authentic MWT-NEI methods cannot provide systematically accurate results. 

\subsection{Virtual blind test}
It should be noted that \emph{real} blind tests such as those carried out in this work were not actually necessary. \emph{Virtual} blind tests would have been sufficient to reach the same conclusion. Indeed, they would be enough to rule out any MWT-NEI method due to the very nature of the MWT problem. To begin with, let's assume that we have a dataset of spectral radiance values, then:
\begin{itemize}
\item we have on one hand a MWT-NEI method (whatever it may be) which is claimed to provides an estimation of the \emph{true} temperature, namely a \emph{unique value for it} (with possibly a confidence interval). The algorithm used to obtain this value is irrelevant here.
\item on the other hand, in a single temperature scenario, there is a continuous infinity of \emph{permitted} solutions, namely exact solutions. The \emph{true} solution is one of them but it is impossible to say which.
\item as a corollary, the MWT-NEI method provides one of these \emph{permitted} solutions (sometimes the solution doublet temperature-emissivity is only approximate) but since no information on the true emissivity was used during the inversion, there is very little chance that the recovered solution corresponds or comes close to the true one. Any MWT-NEI method thus fails almost surely, the success is the same as that with a random draw.
\end{itemize}

However, true blind tests are certainly more striking. 

\subsection{Multiwavelength thermometry (MWT) vs single-wavelength thermometry (SWT)}
By considering the curves of \emph{permitted} solutions in, e.g. Fig. \ref{fig:Case_S_A_B_solutions}, it is clear that the only consequence of removing a wavelength is to possibly increase the range of permitted temperatures $[\hat{T}_L,\,\hat{T}_U]$. By extrapolating down to the case of a single-wavelength measurement, we can deduce that, when emissivity is totally unknown, carrying out a multiwavelength measurement instead of a single-wavelength measurement does not eliminate the impossibility of determining temperature; the only change is that the range of possible values is potentially reduced. Apart from this difference in the range of \emph{permitted temperature}, there is no difference in the potential of SWT and MWT to give an accurate measurement of temperature when nothing is known about the true emissivity spectrum, either in terms of level or shape. 

As stated in Ref.~\citenum{krapez2025commentOE}, MWT-NEI is therefore equivalent, in terms of the dimensionality of the solution space, to the following problem: find $A$ and $B\in[\epsilon_{min},1]$ knowing only their product, which shows that it is a deadlock. This was already the case with SWT-NEI.

We can no longer turn a blind eye to what has been proclaimed since the very beginning of MWT, and which follows from elementary logic: because the problem is underdetermined and has an infinity of solutions, it is necessary to introduce constraints. Obviously, these constraints must be in line with the true emissivity, otherwise the recovered solution is likely to be far from it~\cite{krapez2011measurements,
krapez2019measurements}. Hence MWT without information on emissivity is impossible or with negligible chance of success (except in the trivial cases of VPTR, among which the case where true emissivity spectrum contains very small values and others very close to 1). Methods that only work incidentally are of no interest to the scientific community.

\subsection{Recommended approaches for MWT}
With the spectral radiance data in hand, the first thing to do is to draw a set of \emph{permitted} spectra as, for example, in Fig. \ref{fig:Case_S_A_B_solutions}. Then, think about  how to extract the \emph{true} solution hidden among all the other \emph{permitted} solutions. Have in mind not to be lured by optimization methods based on blind criteria or constraints having no link with the true emissivity spectrum, and therefore cannot help to find the true solution, like e.g. in Ref.~\citenum{zhang2024fast}-\citenum{zhang2025multispectrum}.

One possibility is to capitalize upon the fact that the true emissivity spectrum has this or that shape (e.g. linear, parabolic, exponential, ...), or that for specific wavelengths, emissivity lies within such and such range, or emissivity takes this and that values with given uncertainties.  Obviously, an a priori knowledge of emissivity is mandatory to unblock the situation, e.g. thanks to independent measurements, or, at least as an expedient in the case of metals, by relying on empirical calculations of emissivity via the complex refractive index and resistivity~\citenum{watson2025transient}. It's impossible to escape this a priori knowledge of emissivity. Actually, this statement is valid for the passive MWT configuration considered in this paper; there are other situations where the a priori knowledge of emissivity is not necessary as, for example, with the \emph{spectral smoothness} method, which is a method applied in remote sensing when an hyperspectral infrared sensor is available: the temperature-emissivity separation problem can be solved capitalizing on the fact that the detailed spectral features of the downwelling atmospheric radiance are somehow ‘‘printed’’ in the measured radiance spectrum~\cite{borel1998surface,borel2008error}.

Here, the necessary a priori knowledge about emissivity can be leveraged, at choice:
\begin{itemize}
\item by implementing two-color ratio pyrometry after having checked that the grey assumption is valid~\cite{fagnani2024line}, and if not, the emissivity
ratio at the two wavelengths needs to be determined~\cite{hijazi2011calibrated,musto2016error
,monier2017liquid,watson2025transient}.
\item by implementing the classical LS method after having identified a \emph{suitable} emissivity model~\cite{krapez2011measurements,
krapez2019measurements,qu2020temperature,uman2006fiber,
wen2010assessment,
wen2011temperature,wen2011examination,
yan2020spectral,neupane2022development
,dai2025impact} (notice that \emph{suitable} should be understood as \emph{optimal} in the sense described in §~\ref{subsubsec:Recommended})
\item by applying Bayesian inference~\cite{krapez2019measurements,
suleiman2022bayesian,pierre2022simultaneous,
pierre2024multiple}.
\end{itemize}

In Ref.~\citenum{wen2010assessment,
wen2011temperature,wen2011examination}, MWT was applied on various steel and aluminum alloys and the data were processed by LS with different models assessed sequentially (polynomials, exponential of polynomials, ...). True temperature was measured independently and true emissivity was also available for an a posteriori verification of each model efficiency. Two important observations mere made: LS residuals may have reached low levels, but this did not prevent the calculated temperature from exhibiting both low and high errors. For alloy and model combinations showing a low temperature error, there was also a good match between calculated and experimental emissivity spectra. The authors concluded that
"the closer the generated emissivity spectrum and the measured one, the more accurate is the inferred temperature; i.e., if the emissivity model can represent well the real emissivity behaviors, a more accurate inferred temperature can be achieved."
All this is in line with the interpretation provided in §~\ref{subsubsec:Interpretation}. However, there was no other recommendation on the choice of model than to represent well the real emissivity spectrum. We believe that the criteria given in §~\ref{subsubsec:Recommended} can yield accurate results in a more systematic manner.

Bayesian inference modality involving a priori emissivity values and the LS method involving a parametric model validated with a priori emissivity data have an obvious connection with a third approach which proves to be extremely simple and which consists in the following elementary steps:
\begin{itemize}
\item draw a set of \emph{permitted} spectra like those in Fig.~\ref{fig:Case_S_A_B_solutions}, \ref{fig:Case_R_C_D_solutions},
\item add the a priori emissivity data together with the corresponding uncertainties,
\item find visually and then computationally (for example by mean of weighted LS) which \emph{permitted} spectrum is closest to the a priori data; the associated \emph{permitted} temperature is then the best temperature estimate; its uncertainty can also be easily identified.
\end{itemize}

This third approach achieves a synergistic exploitation of spectral radiance data (converted in terms of a set of permitted solutions) and a priori knowledge of emissivity. It saves the search for an emissivity model, the identification of which requires great care to avoid temperature errors.

\section{CONCLUSION}

A series of blind tests was set up to evaluate objectively the potential of MWT methods claimed to be accurate despite the absence of information on emissivity (MWT-NEI meaning MWT with no emissivity information). These blind tests made it possible to highlight the following points:
\begin{itemize}
\item whatever the scenario (single temperature without reflections, single temperature with reflections, or multi-temperature without reflections), the MWT problem presents a continuous infinity of \emph{permitted} solutions among which the true solution is hidden and impossible to guess without additional information on emissivity (in the multi-temperature scenario with linear dependence of emissivity on temperature, these \emph{permitted} solutions are in fact only quasi-exact solutions; in practice, however, they explain the observed radiance with a RMS error that is much lower than the experimental noise commonly encountered).
\item MWT-NEI methods are based on increasingly sophisticated algorithms which, nevertheless, because they have been given no indication of the true emissivity, provide nothing else than one of these \emph{permitted} solutions (sometimes approximately) without it being possible to say whether it is close to the true solution or not (despite published claims that MWT-NEI methods give accurate results).
\item ultimately, MWT-NEI methods attempt to solve a problem that is unsolvable. The results they provide are misleading because there can be no deterministic link with the true solution.
\end{itemize}


We can now give what appears to be the only correct answer to the blind tests. It is the same for all tests: "\emph{given the spectral radiance data, the range in temperature of the \emph{permitted} solutions is too wide to provide an accurate estimate of the \emph{true} temperature; information on the \emph{true} emissivity is needed to go any further}".

Summing up, it is imperative to include information on emissivity if temperature is to be identified accurately and repeatably. In other words, in MWT, you don't get something for nothing, the same as in single-wave thermometry, after all.

A new paradigm was proposed a few years ago to address the MWT problem; it is based on the systematic use of so-called \emph{permitted} solutions, obtained directly from the measured spectral radiance data. By keeping these \emph{permitted} solutions in mind, many pitfalls can be avoided, such as those revealed by a thorough  literature review, and more efficient search methods for the true solution can be outlined. Thus, assuming spectral emissivity measurements have been carried out beforehand, new criteria have been proposed for an optimal selection of the emissivity model to be used for Least-Squares optimization, with a view to an accurate estimation of the temperature. A method based on a comparative matching between all the \emph{permitted} solutions on the one hand and information that could be obtained on the true emissivity on the other hand was also presented. 

Instead of spending time developing ever more complicated MWT-NEI methods that ultimately serve no purpose, as the present blind tests have vividly demonstrated, it would be preferable to devote efforts to finding the optimum parsimony of emissivity information, i.e. how to solve the MWT problem with the emissivity information that offers the least difficulty to produce.

\acknowledgments 
The author would like to thank the benchmark participants: Ant{\'o}nio Ara{\'u}jo, Shan Gao, Jiashun Luo, Shengxian Shi, Jian Xing, Chunhui Yao, and Kaihua Zhang for providing software applying the Euclidian distance optimization method.



\bibliographystyle{spiebib} 

\end{document}